\documentclass[12pt]{article}

\input epsf
\usepackage{amssymb,amsmath,wrapfig}
\usepackage[matrix,arrow,curve]{xy}
\usepackage{cite}
\usepackage{graphicx}

\makeatletter
\@addtoreset{equation}{section}
\makeatother

\def\rr {{\Bbb R}}
\def\cc {{\Bbb C}}
\def\pp {{\Bbb P}}
\def\zz {{\Bbb Z}}
\def\del {\partial}

\def\del {\partial}

\def\del          {\partial}

\def\sqr#1#2{{\vcenter{\vbox{\hrule height.#2pt
 \hbox{\vrule width.#2pt height#1pt \kern#1pt \vrule width.#2pt}\hrule
 height.#2pt}}}}

\def\beq {\begin{equation}}
\def\bea {\begin{eqnarray}}
\def\eeq {\end{equation}}
\def\eea {\end{eqnarray}}

\begin{document}

	        \begin{titlepage}

	        \begin{center}

	        \vskip .3in \noindent

	        {\Large \bf{Parameter spaces of massive IIA solutions}}

	        \bigskip

	 	 Alessandro Tomasiello and Alberto Zaffaroni\\

	        \bigskip
			 Dipartimento di Fisica, Universit\`a di Milano--Bicocca, I-20126 Milano, Italy\\
            and\\
            INFN, sezione di Milano--Bicocca,
            I-20126 Milano, Italy

	        \vskip .5in
	        {\bf Abstract }
	        \vskip .1in

	        \end{center}

	        \noindent
			We find a new class of ${\cal N}=2$ massive IIA solutions whose internal spaces are $S^2$ fibrations over $S^2\times S^2$. These solutions appear naturally as massive deformations of the type IIA reduction of Sasaki--Einstein manifolds in M--theory, including $Q^{1,1,1}$ and $Y^{p,k}$, and play a role in the AdS$_4$/CFT$_3$  correspondence. We use this example to initiate a systematic study of the parameter space of massive solutions with fluxes. We define and study the natural parameter space of the solutions, which is a certain dense subset of $\rr^3$, whose boundaries correspond to orbifold or conifold singularities. On a codimension--one subset of the parameter space, where the Romans mass vanishes, it is possible to perform a lift to M--theory; extending earlier work, we produce a family $A^{p,q,r}$ of Sasaki--Einstein manifolds with cohomogeneity one and ${\rm SU}(2)\times {\rm SU}(2) \times {\rm U}(1)$ isometry. We also propose a Chern--Simons theory describing the duals of the massless and massive solutions.

	        \vfill
	        \eject


	        \end{titlepage}

\section{Introduction} 
\label{sec:intro}

Understanding the space of solutions of string theory is of paramount importance both for its applications and for more theoretical developments. This was demonstrated long ago  by Calabi--Yau compactifications; the study of their moduli spaces has been enormously rewarding (for a review, see for example \cite{greene-review}). 

The very existence of those moduli has been a motivating factor to the study of flux compactifications. In the best studied classes of examples \cite{grana-polchinski,giddings-kachru-polchinski,kklt}, the internal space is still a Calabi--Yau, whose metric gets distorted by various effects whose ultimate origin is the internal flux. The moduli spaces get then discretized in interesting ways.  

It has been known for a long time, however, that there are more general flux compactifications, whose internal metric is not related to a Calabi--Yau in any way. For examples with no moduli, it is hard to think of an \textit{a priori} way to organize them in a space of parameters, since there is no preexisting moduli space that one is discretizing. 

Recently, some classes of examples have been emerging where there is a ``parameter space'' even though there are no moduli, in an appropriate sense. The massive IIA supergravity solutions on AdS$_4\times \cc\pp^3$ found in \cite{t-cp3}, for example, depend on four parameters; the space of solutions would be $\rr^3\times I$, where $I$ is an interval. Flux quantization discretizes this space in a way reminiscent of the distorted Calabi--Yau solutions; in particular, the discrete set of solutions $D$ in $\rr^3\times I$ has no accumulation points. However, the most interesting parameter is the coordinate $\sigma$ on the interval $I$, which is a ``squashing'' parameter in the internal metric. If one projects $D$ to the interval $I$, one finds that $I$ is densely covered by solutions. In this sense, we have a ``continuous'' space of parameters, $I$, even though there are no (known) moduli. One can also think of this by saying that a certain function of $\sigma$ will have to be rational. 

It is then interesting to study these parameter spaces in other cases. In \cite{koerber-lust-tsimpis} it was found that one of the solutions in \cite{t-cp3} could in fact be generalized; the new parameter space has the shape of an oval in $\rr^2$. In both these massive IIA cases, which have ${\cal N}=1$ supersymmetry, the boundary of the parameter space is made up of solutions with $F_0=0$. (These can then be lifted to AdS$_4\times S^7$ and to AdS$_4\times {\rm SU}(3)/{\rm U}(1)$, respectively.) 

We recently considered \cite{ajtz} a different set of massive solutions on AdS$_4 \times \cc\pp^3$, this time with ${\cal N}=2$ supersymmetry. There is again one parameter that characterizes the internal metric, and its allowed values cover densely an \textit{open} interval $I_0$. As one approaches the endpoints of the interval, one does not find a massless solution, but a singularity in the internal manifold.

In this paper, we consider a more general Ansatz than in \cite{ajtz}. The internal space has generically the topology of an $S^2$ fibration over $S^2\times S^2$. We will find, numerically, a parameter space (see figure \ref{fig:qqp} below) whose boundary will again correspond to singular internal manifolds.

In a sense, this phenomenon actually also appears in K\"ahler moduli spaces of Calabi--Yau manifolds. These look like ``chambers'', with ``walls'' on which two--cycles shrink. In that case, it is known that there is a sense in which one can go ``beyond the wall'' by performing a flop; there is then a notion of extended K\"ahler moduli space, in which one joins several chambers, joined by flops. It is natural to wonder whether similar phenomena might happen here, and whether there are in fact some other string vacua beyond the boundary of our parameter space. A reason to expect that this might happen has to do with the emergence of light branes (as reviewed in section \ref{sec:massive}).

In the interior of the parameter space, along a codimension one subspace, we find cases with $F_0=0$. The local form of these metrics  was actually studied long ago in  \cite[Sec.~4.5]{gauntlett-martelli-sparks-waldram-Apqr} and \cite{chen-lu-pope-vazquezporitz}. We use the analytical form of these massless solutions to study their parameter space, which sits at the intersection of figure \ref{fig:qqp} with a horizontal plane; see figure \ref{fig:ab}. These massless solutions can be lifted to M--theory; the topology of the resulting space $A^{p,q,r}$ depends on three integers $p,q,r$ such that $p,q\le 2 r$  (their ratios being related to the coordinates in figure \ref{fig:ab}). These spaces generalize some previously known Sasaki--Einstein manifolds: for example, $A^{p,p,r}$ is the manifold known as $Y^{r,p}(\cc\pp^1 \times \cc\pp^1)$ \cite{gauntlett-martelli-sparks-waldram-SE7}, whereas $A^{p,p,p}$ is a $\zz_p$ quotient of $Q^{1,1,1}= {\rm SU}(2)^3/{\rm U}(1)^2$ \cite{dauria-fre-vannieuwenhuizen}. 

The full solution with $F_0\ne 0$ is thus a massive deformation of  the type IIA solutions obtained by reducing  particular Sasaki-Einstein manifolds along a supersymmetric direction. The first explicit example of such ${\cal N}=2$ massive deformations  was found in \cite{petrini-zaffaroni} for the particular case of the Sasaki-Einstein manifolds $Y^{p,q}(\mathbb{P}^2)$, by deforming the type IIA reduction of $M^{1,1,1}$.  This result was generalized to manifolds of the type $Y^{p,q}({\rm KE}_4)$ in \cite{lust-tsimpis-singlet-2}, where ${\rm KE}_4$  is a compact four dimensional K\"aler-Einstein manifold. All these massive type IIA solutions are $S^2$ fibration over a ${\rm KE}_4$  
base. The solution considered in this paper  is a $S^2$ fibration over a doubly warped ${\rm KE}_2\times{\rm KE}_2$
base and it contains, as limiting cases, the massive deformations of $Y^{r,p}(\cc\pp^1 \times \cc\pp^1)$ and of quotients of $S^7$.

We will also find a quiver (see figure \ref{fig:tal}) which describes a three--dimensional Chern--Simons--matter theory whose moduli space reproduces the spaces $A^{p,q,r}$, at least when $r\le q\le 2 r$ and $0\le p\le r$. This theory can then be considered a candidate holographic field theory dual to AdS$_4\times A^{p,q,r}$. Following the observation \cite{gaiotto-t,fujita-li-ryu-takayanagi} that $F_0$ corresponds to the sum of Chern--Simons couplings, the quiver theories we consider should also describe the duals to some of the massive solutions.

In section \ref{sec:ansatz}, we will introduce our Ansatz, and study its regularity; we will find a set of easy criteria to decide when solutions will be regular or will have a singularity, and, in the latter case, of which type. In section \ref{sec:massless}, we will review the solutions found in \cite[Sec.~4.5]{gauntlett-martelli-sparks-waldram-Apqr} and \cite{chen-lu-pope-vazquezporitz}, and find their parameter space. In section \ref{sec:top} we will review the topology of the massless solutions, and study their lift $A^{p,q,r}$ to M--theory, including their toric diagram (see figure \ref{fig:td} below). Section \ref{sec:massive} will discuss the parameter spaces of massive solutions. Finally, in section \ref{sec:quiver} we will consider a candidate field theory dual to some of the solutions in this paper.

	
\section{The Ansatz} 
\label{sec:ansatz}

In this section, we will introduce our Ansatz for the metric and fluxes. Locally, this Ansatz is the same as the one considered in \cite{ajtz}; section \ref{sub:local} will review the relevant results in that paper. Our topological setup is more general than the one considered in \cite{ajtz}. We discuss in section \ref{sub:global} under what conditions our Ansatz leads to regular solutions.   

\subsection{Metric and fluxes} 
\label{sub:local}

We are looking for a solution of the form AdS$_4\times M_6$ in IIA string theory. The ten--dimensional metric is fixed by symmetry to be a warped product: 
\begin{equation}\label{eq:46}
	ds^2_{10}= e^{2A} ds^2_{{\rm AdS}_4} + ds^2_6\ .
\end{equation}
Our Ansatz for the internal metric is
\beq
\label{metric6def}
{\rm d} s^2_6 =  \frac{e^{2
B_1(t)}}{4}  {\rm d}s^2_{S^2_1} + \frac{e^{2 B_2(t)}}{4} {\rm
d}s^2_{S^2_2} + \frac{ 1}{8} \, \epsilon^2(t) {\rm d} t^2 +
\frac{1}{64} \Gamma^2(t) ({\rm d} a + A_2-A_1)^2   \ .
\eeq
The range of the coordinate $t$ is the interval $[0,\frac\pi2]$, for reasons we will explain in section \ref{sub:global}. As in \cite{ajtz}, we will refer to $t=0$ as the North pole and $t=\pi/2$ as the South pole. The coordinate $a$ is periodic; its periodicity $\Delta a$ will also be determined in section \ref{sub:global}. Its covariant derivative is defined as
\begin{equation}
	Da \equiv da + A_2 - A_1\ ,
\end{equation}
where $A_i$ are connections with curvatures
\begin{equation}\label{eq:dAJ}
	dA_i = J_i\ ,
\end{equation}
where $J_i$ are the K\"ahler forms of the $S^2_i$. 

The metric (\ref{metric6def}) describes a fibration over the interval $[0,\frac\pi2]$, parameterized by the coordinate $t$. The fibre is itself a U(1) fibration over $S^2\times S^2$. Its Chern class is the element $(4\pi/\Delta a,4\pi/\Delta a)$ of $H^2(S^2\times S^2)$. Thus at this point the spaces $\{t=t_0\}$ could be copies of either $T^{1,1}$ or of any $\zz_k$ quotient thereof. In order for the total six--dimensional space $M_6$ to be compact, the fibre will have to degenerate at the two ends of the interval. We will deal with these issues in section \ref{sub:global}; for the time being, we will deal with the local aspects of (\ref{metric6def}).

The Ansatz (\ref{metric6def}) was also considered in \cite{ajtz}. As we will see later, it includes as particular cases the Fubini--Study metric on $\cc\pp^3$ (which is the internal metric in the ${\cal N}=6$ metric of \cite{nilsson-pope,watamura-2,sorokin-tkach-volkov-11-10}), and some of the metrics in \cite{gauntlett-martelli-sparks-waldram-SE7}. It is natural, then, to wonder whether it contains any more supersymmetric solutions; we will see that it does.

It is convenient to define the combinations
\beq \label{eq:wi}
w_i = 4 e^{2 B_i-2 A} 
\eeq
which control the relative sizes of the two $S^2$'s. As discussed in \cite{ajtz} (see in particular Sec.~5.1 and App.~A there), the supersymmetry equations for ${\cal N}=2$ reduce to three coupled first order ordinary differential equations, for $w_1$, $w_2$ and for a third function $\psi$ which enters in the spinors:
\begin{align}\label{eq:sys}
    \psi' &= \frac{\sin (4 \psi)}{\sin (4 t)}
    \frac{C_{t,\psi}(w_1+w_2) + 2 \cos^2(2t)w_1 w_2 }
    {C_{t,\psi}(w_1+w_2)\cos^2 (2\psi)+2 w_1 w_2}\ ,\nonumber \\
    w_1'&=\frac{4 w_1}{\sin(4t)}
    \frac{C_{t,\psi}(w_1 w_2 - 2\,w_2 - 2\sin^2(2 \psi)w_1) }
    {C_{t,\psi}(w_1+w_2)\cos^2 (2\psi)+2 w_1 w_2}\ ,\\
    w_2'&=\frac{4 w_2}{\sin(4t)}
    \frac{C_{t,\psi}(w_1 w_2 - 2\,w_1 - 2\sin^2(2 \psi)w_2) }
    {C_{t,\psi}(w_1+w_2)\cos^2 (2\psi)+2 w_1 w_2}\ ,\nonumber
\end{align}
where
\begin{equation} \label{eq:Ctpsi}
    C_{t, \psi} \equiv \cos^2(2t)\cos^2(2 \psi)-1.
\end{equation}

All other functions in the metric and the dilaton are algebraically determined in terms of $w_1,w_2,\psi$:
\begin{align}
        \label{eq:eps}
        \epsilon &= \sqrt{2} e^{A} (\cot(\psi) -\tan(\psi) ) \frac{\csc^2 (2t) \, \sin (2\psi) - \cos (2\psi) \, \cot (2t)\,  \psi^\prime}{2 \sqrt{1+ \cot^2 (2t) \, \sin^2 (2\psi) }}\\
        \label{eq:Gamma}
        \Gamma &= 2 e^{A}  \frac{\sin (2 t) +\cos (2 t) \cot (2 t) \sin^2 (2\psi) }{\sqrt{1+ \cot^2 (2t) \, \sin^2 (2\psi) }}\\
        \label{eq:4A}
        e^{4 A} &=  -\frac{4 c}{F_0} \frac{ \sec (2 \psi) \, \tan (2 \psi)}{\sin(4t)} \\
        \label{eq:3Ap}
        e^{3 A -\phi} & = c \sec (2\psi) \sqrt{1+ \cot^2 (2t) \, \sin^2 (2 \psi) } \ .
\end{align}
Here, $c$ is an integration constant, that so far is arbitrary.

The fluxes are completely determined by their internal components $F_0$, $F_2$, $F_4$, $F_6$. These have the general form
\begin{equation}\label{eq:fluxes}
    \begin{split}
        F_2 &= k_2(t) e^{2 B_1} J_1+g_2(t) e^{2 B_2} J_2 + \tilde k_2(t) \frac{i}{2} z\wedge \bar z\ ,\\
        F_4 &= k_4(t) e^{2 B_1+2 B_2} J_1\wedge J_2+\tilde k_4(t) e^{2 B_1} \frac{i}{2} z\wedge \bar z  \wedge J_1 +\tilde g_4(t) e^{2 B_2} \frac{i}{2} z\wedge \bar z  \wedge J_2\ ,\\
        F_6 &= k_6(t) \frac{e^{2B_1+2B_2}}{16}\frac{i}{2} z\wedge \bar z \wedge J_1 \wedge J_2\ ,
    \end{split}
\end{equation}
where $\frac{i}{2} z\wedge \bar z= \frac{\epsilon \Gamma}{16\sqrt{2}} dt\wedge (da+A_2-A_1)$. The coefficients appearing in (\ref{eq:fluxes}) are given by 
\begin{equation}\label{eq:coeffs}
    \begin{split}
        k_2 &= \frac{ c \, e^{-4A}}{ 2 \,w_1 } \frac{\sec (2\psi)}{\cos(2t)} \left ( 2\, C_{t,\psi} +  w_1 \right ),\\
        g_2 &= - \frac{ c\,  e^{-4A}}{ 2 \,w_2 } \frac{\sec (2\psi)}{\cos(2t)}\left ( 2\, C_{t,\psi} +  w_2 \right ),\\
        \tilde k_2 &= 2\,\frac{ c \, e^{-4A}}{ w_1\, w_2} \left ( 2\,C_{t,\psi}(w_1+w_2) + 3\,  w_1\,  w_2 \right ),\\
        k_4 &= -\frac{ c \, e^{-4A}}{4 \, w_1\,  w_2} \frac{ \sin (2\psi) }{\sin(4t) \cos^2(2 \psi)} \left ( 2\,C_{t,\psi}(w_1+w_2) + w_1 \, w_2 \right ),\\
        \tilde k_4 &= \frac{ c \, e^{-4A}}{2 \, w_2 } \frac{\tan (2\psi)}{\sin(2t)} \left ( 2 \,C_{t,\psi} + 3 \, w_2 \right ),\\
        \tilde g_4 &= -\frac{ c \, e^{-4A}}{2 \, w_1 }\frac{\tan (2\psi)}{\sin(2t)} \left ( 2 \,C_{t,\psi} + 3 \, w_1 \right ),\\
        k_6 &= 6\, c \, e^{-4 A}\ .
    \end{split}
\end{equation}
The fluxes (\ref{eq:fluxes}) do satisfy the Bianchi identities, which require that
\beq  \label{eq:tildeF}
\tilde F \equiv  e^{-B} (F_0+F_2+F_4+F_6)
\eeq
is closed. This dictates in particular that $F_0$ is constant.

For $w_1(t)\equiv w_2(t)$ the internal metric is an $S^2$ fibration over $\mathbb{P}^1\times \mathbb{P}^1$ and the 
corresponding Ansatz and solution have been already discussed in  \cite{petrini-zaffaroni,lust-tsimpis-singlet-2}\footnote{\cite{petrini-zaffaroni} discusses the case of an $S^2$ fibration over $\mathbb{P}^2$; however, as originally noticed in \cite{lust-tsimpis-singlet-2},  $\mathbb{P}^2$ can be replaced with any ${\rm KE}_4$, in particular  $\mathbb{P}^1\times \mathbb{P}^1$, without changing any
of the formulae in the paper.}.  The comparison with \cite{petrini-zaffaroni} is particularly easy in our notations:  for $w_1(t)\equiv w_2(t)$  the differential equations (\ref{eq:eps}) reduce to  equations (4.9) in  \cite{petrini-zaffaroni}  after the change of variable $\sqrt{2} \tan( \theta) = \tan{ 2 t}$; the  expressions in the metric and the fluxes also match.  

To summarize, the existence of ${\cal N}=2$ solutions of the form (\ref{eq:46}), (\ref{metric6def}) hinges on the existence of solutions to (\ref{eq:sys}). Solutions to this system are known for $F_0=0$, as we will see in section \ref{sec:massless}; in the $F_0\neq 0$ case, we will study it numerically. 
Let us also remark that there are three trivial symmetries under which (\ref{eq:sys}) is invariant: 
\begin{eqnarray}
	\label{eq:symt} &t \to \frac \pi 2 - t \ ,\qquad &(\psi,w_1,w_2) \to (\psi,w_1,w_2)\ ;\\
	\label{eq:sympsi}&t \to t \ ,\qquad  &(\psi,w_1,w_2) \to (-\psi,w_1,w_2) \ ;\\
	\label{eq:symw}&t \to t \ ,\qquad &(\psi,w_1,w_2) \to (\psi,w_2,w_1)\ .
\end{eqnarray}


\subsection{Regularity and topology} 
\label{sub:global}

The system (\ref{eq:sys}) has a factor $\sin(4t)$ in the denominator. This vanishes in three points within our range $t\in [0,\pi/2]$: $t=0$ (the ``North pole''), $t=\pi/4$ (the ``equator''), $t=\pi/2$ (the ``South pole''). We will begin by studying the system (\ref{eq:sys}) at these three points. 

First of all, notice from (\ref{eq:4A}) that $\psi$ has to go to zero at the poles $t=0,\pi/2$ and equator $t=\pi/4$, if we want $A$ to stay finite there. We can then go on to find the solution as a power series near those points. 

Near the North pole $t=0$, the generic solution is 
\begin{equation}\label{eq:NPgen}
	\begin{split}
        \psi &= \psi_1 t - \frac23 (\psi_1 + 2 \psi_1^3) t^3 + O(t^5)\ ,\\
        w_1 &= w_{10} +
			(2+ 2 \psi_1^2 - w_{10} - w_{10} \psi_1^2) t^2 + O(t^4)\ ,\\
        w_2 &= w_{20} + 
			(2+ 2 \psi_1^2 - w_{20} - w_{20} \psi_1^2) t^2+ O(t^4)\ .
    \end{split}
\end{equation}
Using this expansion in the expression (\ref{metric6def}), the part of the metric containing $t$ and the U(1) direction, up to a constant,  reads
\begin{equation}\label{eq:localgen}
	ds^2= dt^2 + t^2 Da^2\ ,
\end{equation}
up to higher order terms in $t$. The size of the two $S^2$s remains finite. We see that the coordinates $t$ and $a$ combine to give a regular $\rr^2$ if the periodicity of the coordinate $a$ is $\Delta a= 2\pi$. In general, this $\rr^2$ will be fibred over the two $S^2$s, because of the covariant derivative $Da$. For future use, let us also analyze the local behavior of the metric near the poles of the $S^2$s. We use the standard round metric for the spheres, 
$ds^2_{S^2_i}= d\theta_i^2 +\sin^2\theta_i d\phi_i^2$ and  $A_i=-\cos \theta_i d\phi_i$. Near $\theta_i= 0$ and $t=0$, the metric  (\ref{metric6def}) reads
\beq
\label{eq:localgen2}
d s^2_6 = c_1  \left ( d\theta_1^2 +\theta_1^2 d\phi_1^2 \right )  + c_2 \left ( d\theta_2^2 +\theta_2^2 d\phi_2^2 \right )   + c_3 \left (dt^2 + t^2 (da - d\phi_2 + d \phi_1)^2 \right) \ ,
\eeq
with $c_i$ three constants. This is a copy of $\rr^6$, which can be parametrized by three complex coordinates, $z_1=\theta_1e^{i\phi_1}, z_2= \theta_2e^{i\phi_2}$ and $z_3=t e^{i( a -\phi_2+\phi_1)}$. A similar analysis can be performed near the other poles of the two--spheres $S^2$s.

The $O(t^5)$ terms in (\ref{eq:NPgen}), which we have not indicated explicitly, have $w_{10}$ and $w_{20}$ in the denominator. For this reason, the solution (\ref{eq:NPgen}) is not appropriate when either $w_i$ vanishes at $t=0$. If for example $w_2$ vanishes, the solution reads
\begin{equation}\label{eq:NPsp}
	\begin{split}
        \psi &= \psi_1 t - \frac23 (4 \psi_1 +5 \psi_1^3) t^3 + O(t^5)\ ,\\
        w_1 &= w_0 +(4+ 4 \psi_1^2 -2 w_0 +2 w_0 \psi_1^2) t^2 + O(t^4)\ ,\\
        w_2 &= (4+ 4 \psi_1^2) t^2+ O(t^4)\ .
    \end{split}
\end{equation}
This is the case that was considered in \cite{ajtz}. This time, if we use (\ref{eq:NPsp}) in (\ref{metric6def}), the part of the metric involving $t$, $a$ and the shrinking $S^2_2$ approaches, for $t \to 0$,
\begin{equation}\label{eq:localsp}
	dt^2 + \frac14  t^2 \left ( ds^2_{S^2_2} + Da^2\right )\ .
\end{equation}
This is a regular $\rr^4$ (fibred on the non--vanishing sphere $S^2_1$) if we choose the periodicity of the angle $a$ to be $\Delta a = 4\pi$. Notice that this would not be the same choice as the periodicity $\Delta a = 2 \pi$ we encountered after (\ref{eq:localgen}). With that choice, (\ref{eq:localsp}) would instead be the metric on an $\rr^2/\zz_2$ singularity. 

Finally, the case when both $w_i$ vanish is yet another branch to be considered separately -- again because  $w_0$ appears in the denominator of the higher--order terms in (\ref{eq:NPsp}), which we did not show. In this case, the solution is 
\begin{equation}\label{eq:NPcon}
	\begin{split}
        \psi &= \psi_1 t - \frac23 (7 \psi_1 + 8 \psi_1^3) t^3 + O(t^5)\ ,\\
        w_1 &= 6(1+ \psi_1^2) t^2 + O(t^4)\ ,\\
        w_2 &= 6(1+ \psi_1^2) t^2+ O(t^4)\ .
    \end{split}
\end{equation}
If we use this in (\ref{metric6def}), we obtain, again asymptotically for $t\to 0$,
\begin{equation}\label{eq:conifold}
	dt^2+ t^2\left[\frac19 Da^2 + \frac16 \left(ds^2_{S^2_1}+ ds^2_{S^2_2}\right) \right]\ ,
\end{equation}
again up to an overall constant. This is the conifold metric, if the periodicity of the coordinate $a$ is $\Delta a = 4\pi$.

Notice that for all the local metrics we considered -- (\ref{eq:localgen}), (\ref{eq:localsp}) and (\ref{eq:conifold}) -- some part of the metric shrinks at $t=0$. Hence, the range of the parameter $t$ can terminate at $t=0$ without this generating a boundary. That is the reason we chose it as one of the extrema of the allowed range for $t$.

The study of the system near the South pole, $t=\pi/2$, is virtually identical to the study near the North pole we have just seen; so there is no need to treat it here. Again we conclude that the range of the coordinate $t$ can terminate at $t=\pi/2$ without the manifold developing a boundary there. This justifies our choice of $[0,\pi/2]$ as the allowed range for the coordinate $t$.

We now turn to the equator, $t=\pi/4$. We do not want any of the $S^2$s to shrink there. Given this, one can see from (\ref{eq:sys}) that the solution has a singularity (its expansion starts with a term $(t-\pi/4)^{-1}$) unless $w_1=w_2=2$ at $t=\pi/4$. Continuing the perturbative expansion, one finds
\begin{equation}\label{eq:equator}
	\begin{split}
        \psi &= \phi_1 \left(t-\frac\pi4\right) - \frac12 \phi_1 (q_1 + \tilde q_1) \left(t-\frac\pi4\right)^2 + O\left(\left(t-\frac\pi4\right)^3\right)\ ,\\
        w_1 &= 2 + q_1 \left(t-\frac\pi4\right) -\frac14 (32 \phi_1^2 + q_1^2 + q_1 \tilde q_1)\left(t-\frac\pi4\right)^2+ O\left(\left(t-\frac\pi4\right)^3\right)\ ,\\
        w_2 &= 2 + \tilde q_1 \left(t-\frac\pi4\right) -\frac14 (32 \phi_1^2 + \tilde q_1^2 + q_1 \tilde q_1)\left(t-\frac\pi4\right)^2 + O\left(\left(t-\frac\pi4\right)^3\right)\ .
    \end{split}
\end{equation}
Nothing shrinks here; the metric is just $\rr$ times a copy of the fibre, which is $T^{1,1}$ or a quotient thereof, as discussed after equation (\ref{metric6def}). 

Let us summarize the results of this subsection. We found two possible periodicities for the U(1) coordinate $a$ that make sense. 
\begin{itemize}
	\item If the periodicity of $a$ is taken to be $\Delta a = 4 \pi$, we can only consider cases where either one or both $S^2$ shrink at the extrema of the interval $[0,\pi/2]$. If one of the $S^2$ shrinks at a given end, the manifold is regular. At a point where both $S^2$ shrink, the manifold has a conifold singularity.  $\Delta a= 4 \pi$ is the case considered in \cite{ajtz}.
	\item If the periodicity of $a$ is taken to be $\Delta a = 2\pi$, the metric is regular in the generic case when neither of the $S^2$ shrinks at the extrema of the interval $[0,\pi/2]$. If one of the two $S^2$ shrinks at a given end, the manifold has a $\rr^4/\zz_2 \times \rr^2$ singularity there. At a point where both $S^2$ shrink, the manifold has a singularity which is a $\zz_2$ quotient of a conifold singularity. $\Delta a =2\pi$ is the case we will consider in this paper.
\end{itemize}


\section{The massless solutions} 
\label{sec:massless}

In the $F_0=0$ case, the equations (\ref{eq:sys}) can be solved analytically; this solution has locally the form considered in \cite[Sec.~4.5]{gauntlett-martelli-sparks-waldram-Apqr} and \cite{chen-lu-pope-vazquezporitz}. In preparation for the massive case in section \ref{sec:massive}, we will study in detail the global properties of the solutions. The topology of the resulting seven--manifolds will be studied in the next section.

We will begin in section \ref{sub:massless} by taking the $F_0\to 0$ limit of the Ansatz considered in \ref{sub:local}, although this is not strictly necessary. 

\subsection{Massless limit} 
\label{sub:massless}

The system (\ref{eq:sys}) simplifies quite a bit if one takes the limit $F_0\to 0$. Notice first of all that (\ref{eq:4A}) implies that $\psi\to 0$ in this limit. In terms of the $w_i$ defined in (\ref{eq:wi}), we then have
\begin{align}\label{eq:sys0}
	w_1' &= -2 \tan(2t)\frac{w_1 w_2 (w_1-2)}{2 w_1 w_2 - \sin^2(2t) (w_1+w_2)}\\
	w_2' &= -2 \tan(2t)\frac{w_1 w_2 (w_2-2)}{2 w_1 w_2 - \sin^2(2t) (w_1+w_2)}\ . 
\end{align}
The other functions in the metric are then determined by 
\begin{align}
	\frac{\epsilon}{2\sqrt{2}}&= e^A(1- \cot(2t)A') =  \frac{w_1 w_2 e^A}{2 w_1 w_2 - \sin^2(2t) (w_1+w_2)}\ ,\\
	\Gamma&= 2 e^A \sin(2t)\ ,\\
	\label{eq:AeG}
		A' &= \tan(2t)\frac{w_1 w_2 - \sin^2(2t)(w_1+w_2)}{2 w_1 w_2 - \sin^2(2t) (w_1+w_2)}\ ,\\
	\label{eq:3Af}
	e^{3A-\phi} &= c\ .	
\end{align}
(We have manipulated the equations in section \ref{sub:local} so as to make $\psi$ disappear.) We can also determine the fluxes by taking a limit of (\ref{eq:fluxes}) and (\ref{eq:coeffs}). In particular, $H=F_4=0$, and 
\begin{equation}\label{eq:F2}
	F_2  = f J_1 + \tilde f J_2 + g\,dt\wedge Da \ , 
\end{equation}
where the functions $f,\tilde f, g, k$ are given by 
\begin{align}\label{eq:fgen}
	f&= \frac {c\,e^{-2A}}{8 \cos(2t)} (w_1 -2 \sin^2(2t))\ , \\ 
	\label{eq:ftgen}
	\tilde f &
	= -\frac {c\,e^{-2A}}{8 \cos(2t)} (w_2 -2 \sin^2(2t))\ ,\\
	\label{eq:ggen}
	g&= \frac1{8\sqrt{2}}\,\frac{c\, e^{-4A}}{ w_1\, w_2} \left ( -2\sin^2(2t)(w_1+w_2) + 3\,  w_1\,  w_2 \right )\,\epsilon \,\Gamma\ . 
\end{align}
These coefficients satisfy
\begin{equation}\label{eq:dfg}
	\del_t f = - g \ ,\qquad \del_t \tilde f = g \ ,
\end{equation}
as a consequence of the Bianchi identity, that now reads $dF_2=0$.

\subsection{The solutions} 
\label{sub:sol}

In section \ref{sub:massless}, we have seen that the existence of ${\cal N}=2$ solutions in the $F_0=0$ case reduces to a system of two first--order differential equations, (\ref{eq:sys0}). 
In this section, we will solve these equations and we will retrieve the local metric originally found in \cite{gauntlett-martelli-sparks-waldram-Apqr,chen-lu-pope-vazquezporitz} with different methods. 

The first thing to notice is that
$\frac{w_2'}{w_1'}= \frac{(w_2-2)}{(w_1-2)}$. This implies that
\begin{equation}
	\frac{(w_2-2)}{(w_1-2)} = \frac{\tilde q_1}{q_1}
\end{equation}
is a conserved quantity; we have expressed it in terms of the constants $q_1$, $\tilde q_1$ appearing in (\ref{eq:equator}). This means that 
\begin{equation}\label{eq:w2}
	w_2 = 2+ \frac{\tilde q_1}{q_1}(w_1-2)
\end{equation}
is no longer an independent quantity; we have then reduced the existence of solutions to solving a single first--order ODE for $w_1$. As it turns out, by defining
\begin{equation}\label{eq:st}
	s=\cos^2(2t)
\end{equation}
and exchanging the role of the dependent and independent variable, one turns this equation into one which admits an explicit solution, which was found in \cite{gauntlett-martelli-sparks-waldram-Apqr,chen-lu-pope-vazquezporitz}. For ease of comparison with those references, we will introduce
\begin{equation}\label{eq:wr}
	w_1= \rho^2\ ,
\end{equation} 
which is manifestly positive. We should warn the reader that $w_1$  cannot always be thought of as a coordinate (and hence, so cannot $\rho$). This is because it is not always an increasing function of $t$. In the massless case, we will encounter cases where $w_1$ is constant; in the massive case, to be discussed in section \ref{sec:massive}, cases where $w_1$ is non--monotonous are quite common. 

  The solution $s(\rho)$ is given by 
\begin{equation}\label{eq:sr}
	s= \frac{3\rho^2 (\rho^2-2)^2 (2 q_1 + \tilde q_1 (\rho^2-2) )}
	{3 q_1^3 - 2 \tilde q_1 (\rho^2-2)^3-2 q_1 (\rho^2-2)^2(\rho^2+4)}\ .
\end{equation}
 We can now compute all the coefficients in the metric and fluxes. The least trivial step is to integrate the equation (\ref{eq:AeG}) for $A$; fortunately, this leads to a relatively nice expression:
\begin{equation}\label{eq:e2A}
	e^{2A}= c\,\gamma\frac{(2-\rho^2)}{\cos(2t)}= 
	\frac{c\gamma} \rho \sqrt{\frac{3 q_1^3 - 2 \tilde q_1 (\rho^2-2)^3-2 q_1 (\rho^2-2)^2(\rho^2+4)}{3 (2 q_1+ \tilde q_1 (\rho^2-2))}}\ ,
\end{equation}
where we introduced another integration constant $\gamma$.
Rather than giving the explicit form of the other functions $\epsilon$ and $\Gamma$, we will now rewrite the metric using the coordinate $\rho$ defined in (\ref{eq:wr}), in a way directly inspired by \cite{gauntlett-martelli-sparks-waldram-SE7}. We have
\begin{equation}\label{eq:6U}
	ds^2_6= e^{2A}\left[\frac{\rho^2}{16}ds^2_{S^2_1}
	+ \frac{2+(\tilde q_1 / q_1)(\rho^2-2)}{16} ds^2_{S^2_2}+  \frac1U d \rho^2 + q Da^2
	\right]\ ,
\end{equation}
where $U(\rho)$ and $q(\rho)$ are given by
\begin{align}
	\label{eq:U}
	U&= 8\frac{e^{2A}}{\epsilon^2} \left(\frac{\del \rho}{\del t}\right)^2 =
	\frac{(\rho^2-2)^2}{\rho^2}\frac{1-s}s=
	\frac{3 q_1^3 - \tilde q_1 (\rho^2-2)^3 (2+ 3 \rho^2) -8 q_1 (\rho^2-2)^2 (\rho^2+1)}
	{3 \rho^4(2 q_1 + \tilde q_1 (\rho^2-2))}\ , \\ 
	\label{eq:q}
	q&= \frac{e^{-2A}}{64}\Gamma^2= \frac{c^2\gamma^2}{16}e^{-4A}\rho^2 U \ .
\end{align}
Notice also that 
\begin{equation}\label{eq:e4A}
	\frac{e^{4A}}{c^2\gamma^2}=\rho^2 U + (\rho^2-2)^2 \ .
\end{equation}
Later we will also need the functions $f,\tilde f$ in (\ref{eq:F2}), (\ref{eq:fgen}), (\ref{eq:ftgen}): 
\begin{align}\label{eq:fsol}
	f&= -\frac1{8 \gamma}\frac{\rho^2-2 \sin^2(2t)}{\rho^2-2}\ ,\\
	\label{eq:ftsol}
	\tilde f &=\frac1{8 \gamma}\frac{2+(\tilde q_1/q_1)(\rho^2-2)-2 \sin^2(2t)}{\rho^2-2}= - f + \frac{(\tilde q_1/q_1)-1}{8 \gamma} \ ;
\end{align}
we also remark that 
\begin{equation}\label{eq:fU}
	 f= -\frac \gamma {8c^2} \rho^2 e^{-4A} (U + \rho^2-2) \ .
\end{equation}
The relations (\ref{eq:q}), (\ref{eq:e4A}), (\ref{eq:fU}) work in the same way as in \cite{gauntlett-martelli-sparks-waldram-SE7}. 

We will now study the range of $\rho$. First of all, since $w_1=\rho^2$ is the size of the $S^2_1$, we want it not to go to zero, except possibly at the extrema. This means 
\begin{equation}\label{eq:r>0}
	\rho \ge 0 \ .
\end{equation}
Finally, the coefficient of $ds^2_{S^2_2}$ tells us that
\begin{equation}\label{eq:w2pos}
	2+\frac{\tilde q_1}{q_1}(\rho^2-2) \ge 0\ .
\end{equation}
There are, however, more stringent conditions on $\rho$. We have to impose that the functions appearing in (\ref{eq:6U}) are positive. From (\ref{eq:U}), we see that $U\ge 0$ implies
\begin{equation}\label{eq:pr}
	p(\rho)=3 q_1^2 - \frac{\tilde q_1}{q_1} (\rho^2-2)^3 (2+ 3 \rho^2) -8 (\rho^2-2)^2 (\rho^2+1) 
	\ge 0\ .
\end{equation}
Looking at the expression for $U$ in terms of $s$ in (\ref{eq:6U}), we see that imposing $U\ge 0$ implies $0 \le s \le 1$; this takes care of another possible inequality that might have arisen on $\rho$, in view of (\ref{eq:st}). Next, from (\ref{eq:q}) we also see that $U \ge 0$ implies $q\ge 0$. Finally, from (\ref{eq:e4A}) we see that $U \ge 0$ implies that $e^A \ge 0$ too. From (\ref{eq:e2A}) we also see that the sign of $\rho^2-2$ is related to the sign of $\cos(2t)$ -- for a given choice of $\gamma$. For $\gamma>0$, we have
\begin{equation}\label{eq:hemi}
	\rho \le 2 \ \Longleftrightarrow \ \cos(2t) \ge 0 \ ,\qquad
	\rho \ge 2 \ \Longleftrightarrow \ \cos(2t) \le 0\ .
\end{equation} 
Since in this case $\rho$ is an increasing function of $t$, this assignment is appropriate for $q_1>0$. For $q_1<0$, (\ref{eq:hemi}) will have to be reversed. 

Finally, notice that at a zero $\rho_0$ of $U$, $q$ also vanishes, which implies that so does $\Gamma$. Looking back at (\ref{metric6def}), this means that the $S^1$ shrinks. Hence, the zeros of $U$ should correspond to the two extrema $t=0$ and $t=\pi/2$ of the $t$ interval.  
Thus, we can complete (\ref{eq:hemi}):
\begin{equation}\label{eq:hemi2}
	\rho_1 \le \rho \le \sqrt{2}\ \Longleftrightarrow \ 0\le t \le \frac \pi 4  	\ ,\qquad
	\sqrt{2}\le \rho \le \rho_2 \ \Longleftrightarrow \ \frac\pi4\le t \le \frac\pi2 \ .
\end{equation}
where $\rho_1$ and $\rho_2$ are two zeros of $U$. Once again, (\ref{eq:hemi2}) is appropriate for $q_1>0$. For $q_1<0$, this assignment should be reversed; in that case, we will call $\rho_1$ the zero $\ge \sqrt{2}$, and $\rho_2$ the zero $\le \sqrt{2}$. With this convention, for $q_1$ of any sign we will have that  $\rho_1$ corresponds to the North pole $t=0$ and $\rho_2$ corresponds to the South pole $t=\pi/2$.

As a cross--check, recall from (\ref{eq:st}) that $s=\cos^2(2t)$, and from (\ref{eq:U}) that $U=\frac{(\rho^2-2)^2}{\rho^2}\frac{1-s}s$. We see that $U$ vanishes where $s=1$, that is, where $t=0$ or $\pi/2$. It might seem that $U$ also vanishes at $\rho=\sqrt{2}$, but (\ref{eq:sr})  shows that $s$ in the denominator also goes to zero, making $U$ finite. 

To summarize, $U$ should be positive in the interval $[\rho_1,\rho_2]$, and that both $\rho_i$ should satisfy the constraint (\ref{eq:w2pos}). We will now turn to analyzing these inequalities.


\subsection{Allowed range of parameters} 
\label{sub:ineq}

To impose that $U>0$ in an interval, we can just look at the polynomial (\ref{eq:pr}). Since
\begin{equation}
	\del_\rho p = -24 \rho^3 (\rho^2 -2 ) \left(2+ \frac{\tilde q_1}{q_1}(\rho^2-2)\right)\ ,
\end{equation}
the possible extrema of $p$ are
\begin{equation}
	\rho=0 \ ,\qquad \rho= \sqrt{2}\ ,\qquad \rho = \sqrt{2}\sqrt{1-\frac{q_1}{\tilde q_1}}\ .
\end{equation}
From (\ref{eq:6U}) we see that the first value corresponds to a shrinking $S^2_1$, and the third to a shrinking $S^2_2$. As for the second value, it corresponds to $t=\frac\pi 4$, as can be seen from (\ref{eq:hemi}). For future reference, we also compute
\begin{equation}\label{eq:pext}
	p(0)= -32+3 q_1^2 +16 \frac{\tilde q_1}{q_1} \ ,\qquad
	p(\sqrt{2})= 3 q_1^2 \ ,\qquad
	p\left(\sqrt{2}\sqrt{1-\frac{q_1}{\tilde q_1}}\right)= 
	\left(\frac{q_1}{\tilde q_1}\right)^2 \left(-32+3 \tilde q_1^2 +16 \frac{q_1}{\tilde q_1}\right) \ .
\end{equation}
Notice also that 
\begin{equation}\label{eq:p''}
	\del^2_\rho p(\sqrt{2})=-384
\end{equation}
so that $\sqrt{2}$ is always a local maximum.

It will be convenient to divide the analysis in three cases.

\subsubsection{$\tilde q_1/q_1\ge 1$} 
\label{sub:a>1}

In this case, we have
\begin{equation}
	0\le \sqrt{2}\sqrt{1-\frac{q_1}{\tilde q_1}} \le \sqrt{2}\ .
\end{equation}
Notice that the constraint (\ref{eq:w2pos})  in this case imposes
\begin{equation}
	\rho \ge \sqrt{2}\sqrt{1-\frac{q_1}{\tilde q_1}} \ .
\end{equation}
As noticed in (\ref{eq:p''}),  $\sqrt{2}$ is a local maximum; $\sqrt{2}\sqrt{1-\frac{q_1}{\tilde q_1}}$ is a local minimum. Moreover, $p \to -\infty$ for $\rho\to \infty$.

The first zero $\rho_1$ of $U$ (and hence of $p$) should then be located between $\sqrt{2}\sqrt{1-\frac{q_1}{\tilde q_1}}$ and $\sqrt{2}$, so that
\begin{equation}\label{eq:interval-a>1}
	\sqrt{2}\sqrt{1-\frac{q_1}{\tilde q_1}}\le \rho_1 < \sqrt{2} < 
	\rho_2 \ .
\end{equation}
Since the local maximum at $\sqrt{2}$ is automatically positive, as we see from $p(\sqrt{2})$ in (\ref{eq:pext}), for (\ref{eq:interval-a>1}) to happen it is enough to impose that $p$ is negative in $\sqrt{2}\sqrt{1-\frac{q_1}{\tilde q_1}}$. Using (\ref{eq:pext}), this gives
\begin{equation}\label{eq:a>1}
	-32+3 \tilde q_1^2 +16 \frac{q_1}{\tilde q_1} \le 0  \ . \end{equation}


\subsubsection{$0 \le \tilde q_1/q_1< 1$} 
\label{sub:0<a<1}

In this case, $\sqrt{2}\sqrt{1-\frac{q_1}{\tilde q_1}}$ is not real; $p(\rho)$ has only two turning points, a local minimum at $0$ and a local maximum at $\sqrt{2}$; again we have $p \to -\infty$ for $\rho\to \infty$. The constraint (\ref{eq:w2pos})  is automatically guaranteed by $\rho>0$.

The first zero $\rho_1$ of $U$ (and hence of $p$) should then be located between $0$ and $\sqrt{2}$. For this to happen, we should impose that $p$ is negative at the minimum in $0$ (again, it is automatically positive in $\sqrt{2}$). Using (\ref{eq:pext}), this means
\begin{equation}\label{eq:0<a<1}
	-32+3 q_1^2 +16 \frac{\tilde q_1}{q_1} \le 0 \ .
\end{equation}

\subsubsection{$\tilde q_1/q_1 <0$} 
\label{sub:a<0}

In this case, we have
\begin{equation}
	0 < \sqrt{2} < \sqrt{2}\sqrt{1-\frac{q_1}{\tilde q_1}} \ .
\end{equation}
The constraint (\ref{eq:w2pos})  imposes
\begin{equation}
	\rho \le \sqrt{2}\sqrt{1-\frac{q_1}{\tilde q_1}} \ .
\end{equation}
$\sqrt{2}$ is a local maximum, whereas $\sqrt{2}\sqrt{1-\frac{q_1}{\tilde q_1}}$ is a local minimum; this time, $p \to +\infty$ for $\rho\to \infty$.

The zeroes of $p$ should now satisfy
\begin{equation}\label{eq:interval-a<0}
	0 < \rho_1 < \sqrt{2} < \rho_2 < \sqrt{2}\sqrt{1-\frac1 \alpha} \ .
\end{equation}
For this to happen, this time we should impose that $p$ be negative in both $0$ and $\sqrt{2}\sqrt{1-\frac{q_1}{\tilde q_1}}$. Using again (\ref{eq:pext}):
\begin{equation}\label{eq:a<0}
	-32+3 \tilde q_1^2 +16 \frac{q_1}{\tilde q_1} \le 0 \ ,\qquad
	 -32+3 q_1^2 +16 \frac{\tilde q_1}{q_1} \le 0\ .
\end{equation}
The bound given by $p(0)<0$ is stronger for $-1< \frac{\tilde q_1}{q_1} < 0$; the upper bound given by $p(\sqrt{2}\sqrt{1-\frac{q_1}{\tilde q_1}})>0$ is stronger for $\tilde q_1 <-q_1$.


\subsubsection{Summary and limiting values} 
\label{sub:summ}

\begin{figure}[h]
    \centering
        \includegraphics[width=30em]{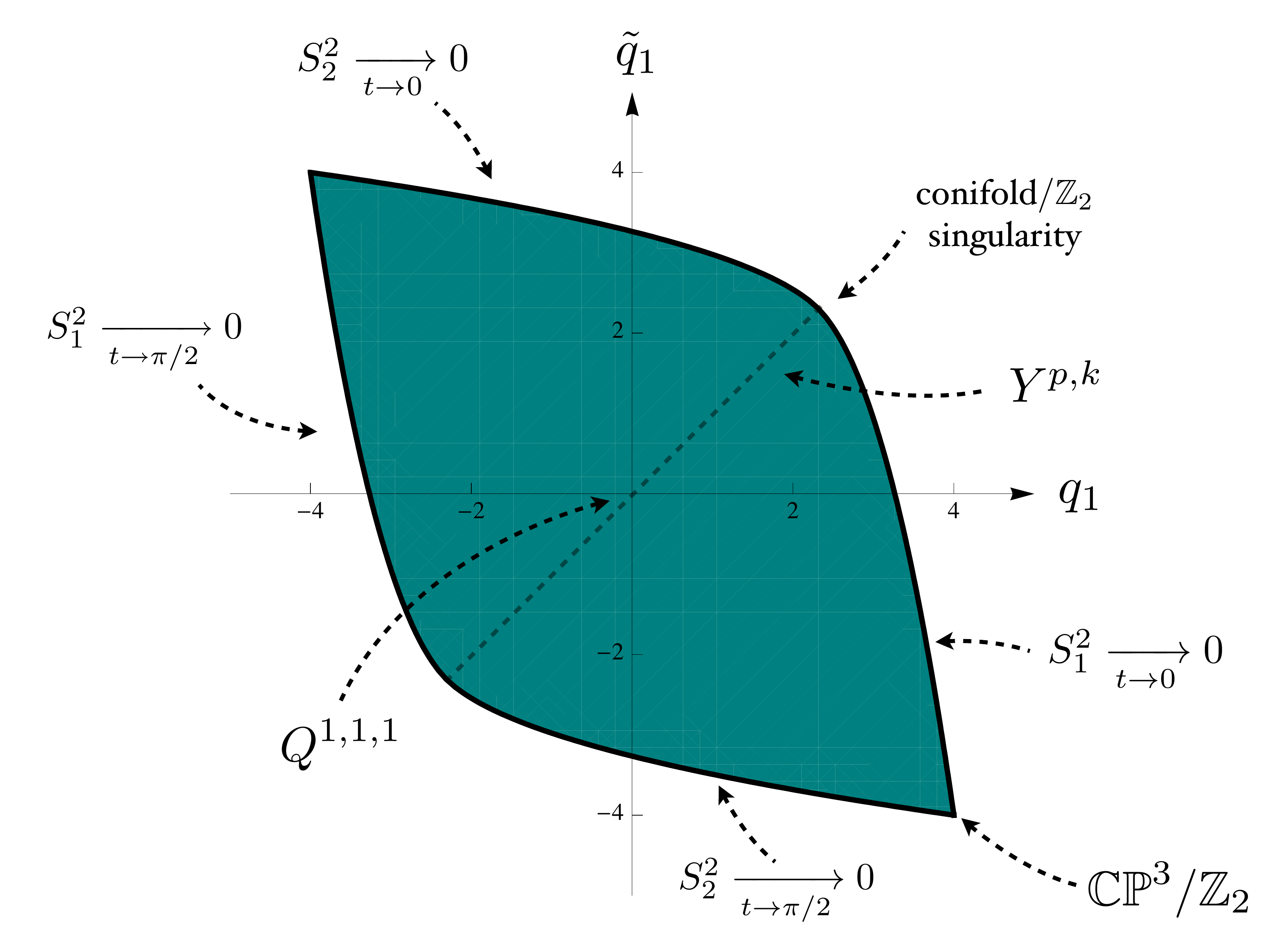}
    \caption{The allowed values for the parameters $(q_1,\tilde q_1)$. Points in the interior of the diagram correspond to non--singular spaces. Generic points on the boundary correspond to manifolds with a $\zz_2$ orbifold singularity.  The points $(\pm 4, \mp 4)$ correspond to spaces with the topology of $\cc\pp^3/\zz_2$. The points $(\pm\frac4{\sqrt{3}},\pm\frac4{\sqrt{3}})$ correspond to spaces with a conifold$/\zz_2$ singularity.}
    \label{fig:ab}
\end{figure}

The bounds we found on $\tilde q_1$ and $q_1$ in (\ref{eq:a>1}), (\ref{eq:0<a<1}) and (\ref{eq:a<0}) are summarized in figure \ref{fig:ab}; we restored the region $q_1<0$, although we only showed our computations for the case $q_1>0$.

This parameter space is symmetric under two reflections: 
\begin{equation}
	(q_1,\tilde q_1) \to (\tilde q_1,q_1) \ ,\qquad
	(q_1,\tilde q_1) \to (-\tilde q_1,-q_1) \ .
\end{equation}
Both are explained by a symmetry of the system: the first by (\ref{eq:symw}), the second by (\ref{eq:symw}) combined with (\ref{eq:symt}). 

There are several loci of interest. Let us first of all consider its boundary. There are four special points there:
\begin{equation}\label{eq:bousp}
	(\pm 4, \mp 4) \ ,\qquad \left(\pm \frac 4{\sqrt{3}}, \pm \frac 4 {\sqrt{3}}\right)\ .
\end{equation}
 These points will be considered later; let us first consider the other, ``generic'' points on the boundary.

Such generic points correspond to a degeneration of one of the two $S^2$. This can be seen from section \ref{sub:ineq}. For example, the bound in (\ref{eq:0<a<1}) comes from demanding that the zero $\rho_1$ of $U$ be non--negative. The limiting case, where the bound (\ref{eq:0<a<1}) is saturated, corresponds to $\rho_1=0$. Since $w_1=\rho^2$ by definition, we see that the sphere $S^2_1$ has collapsed; using (\ref{eq:hemi2}), we see that it has done so for $t\to 0$. As explained at the end of section \ref{sub:global}, when an $S^2$ shrinks we have a $\rr^4/\zz_2 \times \rr^2$ singularity. 
Similar considerations can be applied to other generic components of the boundary. 

Let us now consider the four ``special'' points in (\ref{eq:bousp}). What happens there is that two $S^2$ degenerate. For example, the point $(4,-4)$ corresponds to a case where $S^2_1$ degenerates at the North pole $t=0$, and $S^2_2$ degenerate at the South pole $t=\pi/2$. If the periodicity of the coordinate $a$ were $\Delta a = 4\pi$, the resulting space would have the topology of $\cc\pp^3$, as reviewed in \cite{ajtz} and later in section \ref{sub:base}. In this paper $\Delta a = 2\pi$, and the space is $\cc\pp^3/\zz_2$. 

At the point $(\frac 4{\sqrt{3}},\frac 4{\sqrt{3}})$, again both spheres degenerate -- but this time both at the North pole $t=0$. The local analysis in (\ref{eq:NPcon}), (\ref{eq:conifold}) tells us that this is a $\zz_2$ quotient of a conifold singularity. 

Finally, some loci in figure \ref{fig:ab} correspond to manifolds of note when lifted to M theory. As we will discuss extensively in section \ref{sub:lift}, the $Y^{p,k}\left ( \mathbb{CP}^1\times \mathbb{CP}^1\right )$ manifolds \cite{gauntlett-martelli-sparks-waldram-SE7} correspond to $\tilde q_1=q_1$. The point $\tilde q_1 = q_1 = 0$ is the manifold $Q^{1,1,1}$  \cite{dauria-fre-vannieuwenhuizen}.




\section{Topology and M theory lift} 
\label{sec:top}

In this section, we will study the topology of the six--dimensional base $M_6$ for $F_0=0$, and, after taking care of flux quantization,  the topology of its seven--dimensional manifold lift to M--theory (which will be a Sasaki--Einstein space). We will also study the toric structure of the Calabi--Yau cone over the Sasaki--Einstein space.

\subsection{The six--dimensional base} 
\label{sub:base}

For generic values of $(q_1, \tilde q_1)$ within the allowed region in figure \ref{fig:ab}, the topology of $M_6$ is the same as in \cite{gauntlett-martelli-sparks-waldram-SE7}. Namely, we know that the $S^1$ described by the coordinate $a$ is fibred over the two $S^2$s, and that the Chern class of the fibration is the element $(2,2)$ in $H^2(S^2 \times S^2)$. The associated $\rr^2$ bundle over $S^2\times S^2$ is then the anticanonical bundle $K^*={\cal O}_{\cc\pp^1 \times \cc\pp^1}(2,2)$. The associated $S^2$ bundle is then
\begin{equation}\label{eq:topgen}
	\pp (K^* \oplus {\cal O})\ ,
\end{equation}
which can also be written as $\pp({\cal O} \oplus K)$, as in \cite{gauntlett-martelli-sparks-waldram-SE7}.

Values of $(q_1, \tilde q_1)$ on the boundary of the diagram in figure \ref{fig:ab} lead to different topologies. This is already clear from the fact that $M_6$ develops $\zz_2$ singularities there, as we found in section \ref{sub:summ}. Let us first look at a ``generic'' point on the boundary, one that differs from the four ``corners'' (\ref{eq:bousp}). All these generic points have the same global topology; they only differ by which of the two $S^2$s degenerates at which end of the interval $t\in [0,\pi/2]$. (We already know that points $q_1=\pm \tilde q_1$ are special, and we will deal with them later.) We thus have to do with a $T^{2,2}=T^{1,1}/\zz_2$ fibration over an interval, such that one of the $S^2$ in the base of $T^{2,2}$, as well as the $S^1$ in its fibre, degenerate at one end; at the other end, only the $S^1$ degenerates. 

To decide what this topology is, it is useful to take a step back and think about a simpler case, where we take away from the picture one of the $S^2$s, the one that never shrinks. This means considering again a  fibration over an interval, but one whose fibre is itself a $S^1$ fibration over an $S^2$; at one end, both $S^1$ and $S^2$ shrink, whereas at the other end the $S^1$ alone shrinks. If the $S^1$ fibration over $S^2$ is an $S^3$ (and not one of its quotients), the total four--dimensional space we have obtained has the topology of a $\cc\pp^2$. 

One way to see this is to consider $S^5$ as a $S^1 \times S^3$ fibration over an interval; the $S^1$ shrinks at one end, the $S^3$ at the other end. Let us reduce this $S^5$ along the diagonal U(1) that mixes the first $S^1$ and the Hopf $S^1\subset S^3$. We know that the reduction of $S^5$ along a never--vanishing vector is $\cc\pp^2$. This gives rise to a $S^1 \times S^3/S^1 = S^3$ fibration over the interval; at one end of the interval this degenerates to $S^3/S^1 = S^2$, at the other end to $S^1/S^1=$ a point. This proves the claim in the previous paragraph about fibrating $\cc\pp^2$ over an interval. Another way to see this is to consider $\cc\pp^2$ as a toric manifold. Its toric polytope is a triangle; we can then see $\cc\pp^2$ as a $T^2$ fibration over this triangle, where the $T^2$ degenerates on the boundary according to a pattern captured by the slopes of its sides. We can now fibrate this triangle over one of its sides, so that the fibre is a segment that shrinks to a point at one end. If we add the $T^2$ fibre, this reproduces again the claim in the previous paragraph.  

Let us now go back to our six--dimensional manifold corresponding to generic points on the boundary of figure \ref{fig:ab}. Recall that only one of the $S^2$ shrinks at one end of the interval; to fix ideas, let us consider the upper branch of the boundary, between $(-4,4)$ and $(4/\sqrt{3},4/\sqrt{3})$, where $S^2_2$ shrinks at $t=0$. The other sphere, $S^2_1$, does not shrink at either end.  First of all, the periodicity of $a$ is $2\pi$, and not $4\pi$ as it should be for a Hopf fibration. So the $S^1$ parameterized by $a$, the sphere $S^2_2$ and the interval parameterized by $t$ describe together a $\cc\pp^2/\zz_2$, rather than a $\cc\pp^2$ as in the previous two paragraphs. Moreover, we also have the other sphere $S^2_1$, that does not shrink at either end. The total space is not simply a product $S^2\times\cc\pp^2/\zz_2$, because $a$ is fibred over $S^2_1$ too. In fact we have a fibration
\begin{equation}
	\xymatrix{ 
	\cc\pp^2/\zz_2 \ \ar@{^{(}->}[r] &M_6\ar[d]\\
	& S^2}\ .
\end{equation}
The same considerations apply to all other generic points on the boundary. 

We now turn to the special points on the boundary, (\ref{eq:bousp}). The cases $(4/\sqrt{3},4/\sqrt{3})$ correspond to a six--dimensional space with a singularity which is a $\zz_2$ quotient of a conifold singularity. It can be viewed as a $\cc\pp^1$ bundle over $\cc\pp^1 \times \cc\pp^1$, with one section blown down to a point. 

The cases $(\pm 4, \mp 4)$ are $\cc\pp^3/\zz_2$, as was already remarked in section \ref{sub:summ}. The fact that $\cc\pp^3$ can be seen as a $T^{1,1}$ fibration over an interval was already noticed in \cite{cvetic-lu-pope-cpn}, and used in \cite{gaiotto-t2,ajtz}. It can be seen by reducing along a U(1) the more standard realization of $S^7$ as a $S^3 \times S^3$ fibration over an interval; this goes along similar lines to the fibration for $\cc\pp^2$ we discussed earlier in this subsection. We can also use the toric picture of $\cc\pp^3$ as a $T^3$ fibration over its toric polytope, a tetrahedron. We can slice that tetrahedron so as to have a fibration over an interval whose generic fibres is a square, that degenerates to a segment over each end. Including the $T^3$ fibres leads again to the claimed fibration of $\cc\pp^3$ over an interval.


\subsection{Flux quantization} 
\label{sub:periods}

So far we have not paid any attention to flux quantization. We are going to take care of it in this section. 

First of all, let us count the number of constants that parameterize our solutions. We have the constant $c$ in (\ref{eq:3Af}), the integration constant $\gamma$ we introduced in (\ref{eq:e2A}), and the two parameters $q_1$ and $\tilde q_1$ we see in figure \ref{fig:ab}, for a total of four parameters. $c$ and $\gamma$  can be interpreted by defining $g_s \equiv e^\phi|_{t=0}$ and $L \equiv e^A|_{t=0}$ to be the value of the string coupling and AdS$_4$ radius at one particular point in the internal manifold. From (\ref{eq:3Af}) and (\ref{eq:e2A}) we then have
\begin{equation}
	c= \frac{L^3}{g_s} \ ,\qquad \gamma= \frac{g_s}{2L}\ .
\end{equation}

Let us now count the flux quantization conditions. 
For values of $(q_1, \tilde q_1)$ that are in the interior of the allowed region in figure \ref{fig:ab}, we have seen that $M_6$ has the topology (\ref{eq:topgen}). In particular, there are three two--cycles $B_i$, $i=1,2,3$. Thus, we will have four quantization conditions: 
\begin{equation}
	\Pi_i\equiv\int_{B_i} F_2 = n_{2,i} (2\pi l_s)\ ,\qquad 
	\int_{M_6} F_6=n_6 (2\pi l_s)^5 \ .
\end{equation} 
We thus will have four conditions on four real parameters; we expect to find solutions at least for a certain range of the flux integers. We will now determine that range. 

First of all, let us consider the quantization law for $\int_{M_6} F_6$ and for one of the two--cycles, say $\Pi_3=\int_{B_3} F_2$. These integrals will result in expressions of the form $\frac{L^5}{g_s} f_6(q_1, \tilde q_1)$ and $\frac{L}{g_s} f_{2,3}(q_1, \tilde q_1)$, respectively. We can invert these to produce expressions for $L$ and $g_s$ in terms of $f_6$, $f_{2,3}$ and of the corresponding flux integers $n_6$, $n_{2,3}$. 

Things are more interesting when we consider the two remaining quantization conditions, $\Pi_i=\int_{B_i} F_2$, $i=1,2$. These will also produce expressions of the form $\frac{L}{g_s} f_{2,i}(q_1, \tilde q_1)$. We have already determined $L$ and $g_s$; by considering the quotients $\Pi_3 / \Pi_i$, $i=1,2$, we will obtain two conditions on $q_1$ and $\tilde q_1$ that are completely disentangled from the conditions on $L$ and $g_s$. We are thus reduced to imposing that
\begin{equation}\label{eq:Q}
	\frac{\Pi_3}{\Pi_i} \in {\Bbb Q}\ .
\end{equation}
From the point of view of M--theory, one might think that the right condition to impose is a stronger one, namely that the $\Pi_i \in \zz$, $i=1,2,3$. However, we are free to choose the periodicity of the eleventh direction $\psi$ as we like, and this accounts for the weaker condition (\ref{eq:Q}). 
 
We now turn to imposing (\ref{eq:Q}) explicitly. As in \cite{gauntlett-martelli-sparks-waldram-SE7}, we can take the homology representatives as follows. $B_1$ and $B_2$ will be respectively  the first sphere $S^2_1$ and minus the second sphere $S^2_2$, both at $t=\pi/2$. $B_3$ will be minus the two--sphere spanned by the coordinates $(t,a)$. All three periods can be computed in terms of the two zeros $\rho_1$ and $\rho_2$ of $U$ (which correspond to $t=0$ and $t=\pi/2$ respectively). We can recall that the functions $f$, $\tilde f$ in (\ref{eq:F2}) are given by (\ref{eq:fsol}) and (\ref{eq:ftsol}); for the function $g$, one can use (\ref{eq:dfg}). We get:
\begin{align}
	\Pi_1&\equiv\int_{B_1} F_2 = 4\pi f|_{t=\pi/2}= -\frac \pi {2 \gamma} \frac{\rho_2^2}{\rho_2^2-2} \ , \\ 
	\Pi_2&\equiv\int_{B_2} F_2 = -4\pi \tilde f|_{t=\pi/2}= \Pi_1 - 
	\frac \pi{2 \gamma}\left(\frac{\tilde q_1}{q_1}-1\right) =
	-\frac \pi {2 \gamma} \frac{2+(\tilde q_1/q_1)(\rho_2^2-2)}{\rho_2^2-2} \ , \\
	\begin{split}
			\Pi_3&\equiv\int_{B_3} F_2 = 2\pi (f|_{t=\pi/2}- f|_{t=0})=-\frac \pi {4 \gamma} \left(\frac{\rho_2^2}{\rho_2^2-2}- \frac{\rho_1^2}{\rho_1^2-2}\right) \\
			&=- \frac \pi {4 \gamma}\left(\frac{2+(\tilde q_1/q_1)(\rho_2^2-2)}{\rho_2^2-2} - \frac{2+(\tilde q_1/q_1)(\rho_1^2-2)}{\rho_1^2-2}\right)\ .
	\end{split}
\end{align}
Condition (\ref{eq:Q})  thus reduces to
\begin{equation}
	\label{eq:pi3/pii}
	\frac{\Pi_3}{\Pi_1}= \frac12\left(1-\frac{\rho_1^2 (\rho_2^2-2)}{\rho_2^2 (\rho_1^2-2)}\right) \equiv \frac rp\ ,\qquad
	\frac{\Pi_3}{\Pi_2}=\frac12\left(1-\frac{(2+ (\tilde q_1/q_1)(\rho_1^2-2)) (\rho_2^2-2)}
	{(2+(\tilde q_1/q_1)(\rho_2^2-2)) (\rho_1^2-2)}\right) \equiv \frac rq \ ,
\end{equation}
where $p,q,r$ are integers. Notice that these conditions impose that two functions of the parameters should be rational, not integer. Thus, the parameter space in figure \ref{fig:ab} is discretized, but \textit{densely}. 

Although it is not manifest in these formulas, $\Pi_1$ and $\Pi_2$ are exchanged by $q_1 \leftrightarrow \tilde q_1$, since this operation corresponds to the symmetry (\ref{eq:symw}) that exchanges the two spheres.

We can study the functions $\Pi_3/\Pi_i$ on our space of allowed $(q_1,\tilde q_1)$ in figure \ref{fig:ab}. It is easiest to do so at the boundary of that space; one can then see that this is where the extreme values are reached. For $\Pi_3/\Pi_1$, we read from figure \ref{fig:ab} that on the rightmost boundary in figure \ref{fig:ab}, between $(4/\sqrt{3},4/\sqrt{3})$ and $(4,-4)$, the sphere $S^2_1$ shrinks at $t\to 0$; so $\rho_1^2$ goes to 0 there, and $\Pi_3/\Pi_1=1/2$. On the leftmost boundary, on the other hand, the sphere $S^2_1$ shrinks at $t\to \pi/2$, so $\rho_2^2$ goes to 0 there, and $\Pi_3/\Pi_1\to \infty$. The image of $\Pi_3/\Pi_1$ is  $[1/2,\infty]$. 

The other ratio $\Pi_3/\Pi_2$ has a similar behavior, as can be inferred by its being exchanged with $\Pi_3/\Pi_1$ under $q_1\leftrightarrow \tilde q_1$. We see from figure \ref{fig:ab} that on the upper boundary in figure \ref{fig:ab}, between $(-4,4)$ and $(4/\sqrt{3},4/\sqrt{3})$, the sphere $S^2_2$ shrinks at $t\to 0$; so $2+ (\tilde q_1/q_1)(\rho_1^2-2)$ goes to 0 there, and $\Pi_3/\Pi_2=1/2$. On the lower boundary, on the other hand, the sphere $S^2_2$ shrinks at $t\to \pi/2$, so $2+ (\tilde q_1/q_1)(\rho_2^2-2)$ goes to 0 there, and $\Pi_3/\Pi_2\to \infty$. The image of $\Pi_3/\Pi_2$ is again  $[1/2,\infty]$. 

These results on $\Pi_3/\Pi_i$ give us
\begin{equation}\label{eq:rangeZ}
	0 \le  p \le 2r  \ ,\qquad
	 o \le q \le 2r \ . 
\end{equation}
or an isomorphic manifold with $(p\rightarrow -p, q\rightarrow -q,r\rightarrow -r)$.
For any integer within this range, the flux quantization conditions can be satisfied. 

\subsection{The M--theory lift} 
\label{sub:lift}

We have now shown that one finds IIA solutions parameterized by three integers $p,q,r$ (as well as the flux integer $n_6$).  Since these solutions are massless, we can lift them to M--theory solutions of the form AdS$_4\times M_7$. Since $F_2$ will become part of the eleven--dimensional geometry, the only flux present will be internal $G_7$, and its dual $G_4$ along the spacetime. This is a Freund--Rubin compactification; since we have  ${\cal N}=2$ supersymmetry, $M_7$ will be a Sasaki--Einstein manifold (namely, one whose cone is a Calabi--Yau fourfold). 
We will call these seven--manifolds $A^{p,q,r}$. The local form of these  metrics already appeared in \cite{gauntlett-martelli-sparks-waldram-Apqr,chen-lu-pope-vazquezporitz} and, in coordinates where the $SU(2)\times SU(2)$ symmetry is not manifest, in \cite{lu-pope-vazquezporitz2}.  

We can study the local geometry of these spaces. In general, the relation between the seven--dimensional metric in M--theory and the six--dimensional internal metric in IIA reads
\begin{equation}\label{eq:76}
	ds^2_7  = e^{-\frac23 \phi} ds^2_6 + e^{\frac43 \phi} (d \psi + C_1)^2\ ,
\end{equation}
where $\phi$ is the dilaton, $\psi$ is the periodic eleventh direction, and $C_1$ is the RR one--form potential, defined such that 
\begin{equation}\label{eq:dCF}
	dC_1 = F_2\ .
\end{equation}
We know already the expressions for $\phi$ and $F_2$ in our solutions. A possible choice for $C_1$ that satisfies (\ref{eq:dCF}) is 
\begin{equation}
	C_1 = \left(-f +\frac{\tilde q_1-  q_1}{16  q_1 \gamma}\right) Da + \frac{\tilde q_1- q_1}{16  q_1\gamma}(A_1 + A_2)\ .
\end{equation}
We know that the periods of $F_2$ are rationally related and we can write them as $\Pi_1= 2\pi l p, \Pi_2= 2\pi l q,\Pi_3= 2 \pi l r$
with $l\in \mathbb{R}$. The periodicity of $\psi$ is then $2\pi l$.

If we now apply (\ref{eq:76}) to our metric (\ref{eq:6U}), recalling the relation between $\phi$ and $3A$ in (\ref{eq:3Af}) we see that the factor of $e^{-\frac23 \phi}$ cancels the overall factor of $e^{2A}$ in (\ref{eq:6U}), and we end up with
\begin{equation}
	ds^2_7  = \frac{w_1}{16}ds^2_{S^2_1}
	+ \frac{w_2}{16 }  ds^2_{S^2_2}+  \frac1U d \rho^2 + q Da^2 + \gamma^2[\rho^2 U + (\rho^2-2)^2 ] (d \psi + C_1)^2\ ,
\end{equation}
up to an overall factor of $c^{-2/3}$. Here, we have also used (\ref{eq:e4A}). We recall that $w_1=\rho^2$ and $w_2=2+\frac{\tilde q_1}{q_1}( \rho^2-2)$. If we now define the variable
\begin{equation}
	\tau = \psi+ \frac{\tilde q_1- q_1}{16  q_1 \gamma} a\ ,
\end{equation}
we see that the terms involving $(da)^2$ simplify drastically, giving the constant $1/16$. We get
\begin{align}\label{eq:localfibr}
	ds^2_7 &= d\tilde s^2_6 +\left(\frac 14 d a + \sigma\right)^2\ ,\\
	\sigma &= \frac18 (-w_1 A_1 + w_2 A_2) + \gamma (\rho^2-2) d \tau\ ,\\
	d\tilde s^2_6 &=\frac{w_1}{16}ds^2_{S^2_1}
	+ \frac{w_2}{16} ds^2_{S^2_2}+  \frac{ d \rho^2}{U}+ 
	 \rho^2 U \left(\gamma d \tau -\frac1{8  q_1}( q_1A_1 - \tilde  q_1 A_2)\right)^2\ .
\end{align}

We now show that the seven dimensional metric is  Sasaki--Einstein\footnote{We refer to \cite{MS} for a useful  review of  notions of Sasaki and toric geometry that are used in the following.}. In (\ref{eq:localfibr}) the metric is written as a fibration over a six--dimensional local base. If we define a local K\"ahler  two--form $J$ and a local holomorphic  $(3,0)$ form  $\Omega$ on the six-dimensional base: 
\begin{align}
J &= \frac{i}{2} \left ( z_1 \wedge \bar z_1 +z_2 \wedge \bar z_2+ z_3 \wedge \bar z_3 \right )\ ,\\
\Omega &=  z_1\wedge z_2\wedge z_3 \ ,
\end{align} 
where
\begin{align}
z_1 &= \frac{\sqrt{w_1}}{4} \left ( d\theta_1 - i \sin\theta_1 d\phi_1 \right ) e^{i \frac{a}{2}}\ , \\
z_2 &=  \frac{ \sqrt{w_2}}{4} \left ( d\theta_2 + i \sin\theta_2 d\phi_2 \right ) e^{i \frac{a}{2}}\ , \\ 
z_3 &= \frac{d\rho}{\sqrt{U}} + i \rho \sqrt{U} \left(\gamma d \tau -\frac1{8  q_1}( q_1A_1 - \tilde  q_1 A_2)\right) \ ,
\end{align}
the Sasaki--Einstein conditions are equivalent to the following differential constraints:
\begin{equation}
d\sigma=2 J\, , \qquad\qquad\qquad d\Omega=4 i \left ( \frac{1}{4} da +\sigma\right ) \wedge \Omega\ ,
\end{equation} 
which are easy to verify. The metric  $d\tilde s^2_6$ is K\"ahler--Einstein, but, unlike $d s^2_6$, it is in general only locally defined.   

The local form of these  Sasaki--Einstein metrics already appeared in \cite{gauntlett-martelli-sparks-waldram-Apqr,chen-lu-pope-vazquezporitz}. The global structure has been previously discussed only for the  special case of $p=q$ ($\tilde q_1=q_1\, , 
w_1=w_2$) where we obtain the metric of $Y^{r,p}\left ( \cc\pp^2\times \cc\pp^1\right )$ \cite{gauntlett-martelli-sparks-waldram-SE7}. We have now shown that there is a three integer parameters family of Sasaki--Einstein metrics
$A^{p,q,r}$  with $SU(2)\times SU(2)$ isometry that reduces to $Y^{r,p}\left ( \cc\pp^2\times \cc\pp^1\right )$ in the case $p=q$.

\subsubsection{The toric diagram} 

Given any Sasaki--Einstein metric, we can construct the Calabi--Yau cone
\begin{equation}
ds_8^2 = dr^2 + r^2 ds^2_7 
\end{equation}
with K\"ahler and holomorphic $(4,0)$ forms:  
\begin{equation}
J_4 = d\left [ \frac{r^2}{2}\left ( \frac{1}{4} da+ \sigma\right )\right ]\, , \qquad\qquad \Omega_4 = r^3 \Omega\wedge \left [ dr+ i r\left ( \frac{1}{4} da+\sigma\right )\right ]\, ,
\end{equation}
which obviously satisfy $dJ_4=d\Omega_4=0$. In our particular case, the Calabi-Yau cone is a toric manifold. 
We know that we have four angular variables $\psi^\prime \equiv\psi/l\, ,a\, ,\phi_1\, ,\phi_2$, all with periodicity $2\pi$. They define a $\mathbb{T}^4$ action on the manifold. However,  we need to pay some attention  to the 
choice of a basis in $\mathbb{T}^4$. We need four vectors $t_i=\partial /\partial\phi_i$ which define closed
orbits of length $2\pi$;  in other words, $e^{2\pi i t_i}$ should be the identity transformation on the manifold. We also need to choose them so that the action of this $\Bbb{T}^4$ is effective: namely, so that none of its elements acts like the identity on the manifold.  
A possible choice is $\phi_1\, ,\phi_2\, ,\phi_3\equiv - a\, , \phi_4 \equiv  \psi^\prime -\frac{q+p}{4} a +\frac{p}{2} \phi_1+\frac{q}{2} \phi_2$. The corresponding vectors are 
\bea
t_1 &=& \frac{\partial}{\partial \phi_1} - \frac{p}{2}  \frac{\partial}{\partial \psi^\prime}\ , \nonumber\\
t_2 &=& \frac{\partial}{\partial \phi_2} - \frac{q}{2}  \frac{\partial}{\partial \psi^\prime}\ , \nonumber\\
t_3 &=& - \frac{\partial}{\partial a} - \frac{q+p}{4}  \frac{\partial}{\partial \psi^\prime}\ , \nonumber\\
t_4 &=& \frac{\partial}{\partial \psi^\prime} \ .
\label{vectors}
\eea 
The only subtle point here is the shift in $\psi^\prime$ that we will now explain. We can find a good basis for the vectors by looking at the loci where they degenerate.  The  $\mathbb{T}^4$ action
degenerates to an action of $\mathbb{T}$ on the eight complex lines $\{ \rho=\rho_1,\rho_2\, ;\theta_1=0,\pi \, ;\theta_2=0,\pi \}$. Near each of these lines, the seven--dimensional Sasaki--Einstein manifold should look like
 $\mathbb{R}^6\times S^1$.   For example, we can look at the locus  $\rho=\rho_2\, ,\theta_1=\pi \, , \theta_2=0$.
 On the six--dimensional base $ds_6^2$ in (\ref{eq:76}), this is the point corresponding to the South pole $t=\pi/2$ on the fiber and to a choice of poles on $S_1^2$ and $S_2^2$.  We known from the analysis in section \ref{sub:global} (see in particular the discussion around (\ref{eq:localgen2})), that the metric is locally $\mathbb{R}^6$, and $\phi_1^\prime=\phi_1\, ,\phi_2^\prime=\phi_2\, ,\phi_3^\prime \equiv - (a -\phi_1-\phi_2)$ is a good choice of angular coordinates near this point.  On the other hand, we know from equation (\ref{eq:76}) that $\psi^\prime$ appears in the combination
 \begin{equation}
 \left ( d\psi + \frac{\tilde f -f}{2} da + f d\phi_1 -\tilde f  d\phi_2\right )^2\Big |_{t=\pi/2}= l^2  \left ( d\psi^\prime - \frac{q+p}{4} da + \frac{p}{2}  d\phi_1 +\frac{q}{2}  d\phi_2\right )^2 \ .
\end{equation} 
We see that  $\phi_4^\prime \equiv  \psi^\prime -\frac{q+p}{4} a +\frac{p}{2} \phi_1+\frac{q}{2} \phi_2$ is a good coordinate for $S^1$. The  vectors $t_i^\prime=\partial /\partial\phi_i^\prime$ do have closed orbits of period $2\pi$ near the degeneration locus and can be smoothly extended to the entire manifold.  It is easy to check that the analysis near the other fixed lines gives  $SL(4,\mathbb{Z})$ equivalent basis of vectors.
In order to compare with the results in  \cite{martelli-sparks-sasaki7}, we use the  vectors $t_i$ in (\ref{vectors}) that are related to the $t_i^\prime$ by  the $SL(4,\mathbb{Z})$ transformation $t_1^\prime = t_1-t_3,\, t_2^\prime = t_2-t_3,\, t_3^\prime=t_3,\, t_4^\prime = t_4$.

The four vectors $t_i=\partial /\partial\phi_i$ have closed orbits and  define an effective Hamiltonian $\mathbb{T}^4$ action on the cone over the Sasaki--Einstein manifold with respect to the symplectic form
 \begin{equation}
J_4 = d\left [ \frac{r^2  w_1}{16} \cos\theta_1 d\phi_1- \frac{r^2  w_2}{16} \cos\theta_2 d\phi_2 +\frac{r^2}{2} (\rho^2-2)  \gamma d\psi
 +   \frac{q_1 (2 +u^2)+\tilde q_1 (u^2-2)}{32 q_1} da \right ]  \ ,
 \end{equation}
which can be also written as $J_4= \sum_{i=1}^4 d \mu_i \wedge d\phi_i$.  According to general results of symplectic geometry, the image of the four Hamiltonians (momentum maps) $\mu_i$
in $\mathbb{R}^4$ is a convex rational polyhedron.  And in fact, we see that  the images of the eight lines $\{ \rho=\rho_1,\rho_2\, ;\theta_1=0,\pi \, ; \theta_2=0,\pi \}$ lie on the directions specified by the integer vectors 
\begin{figure}[h]
    \centering
        \includegraphics[width=13em]{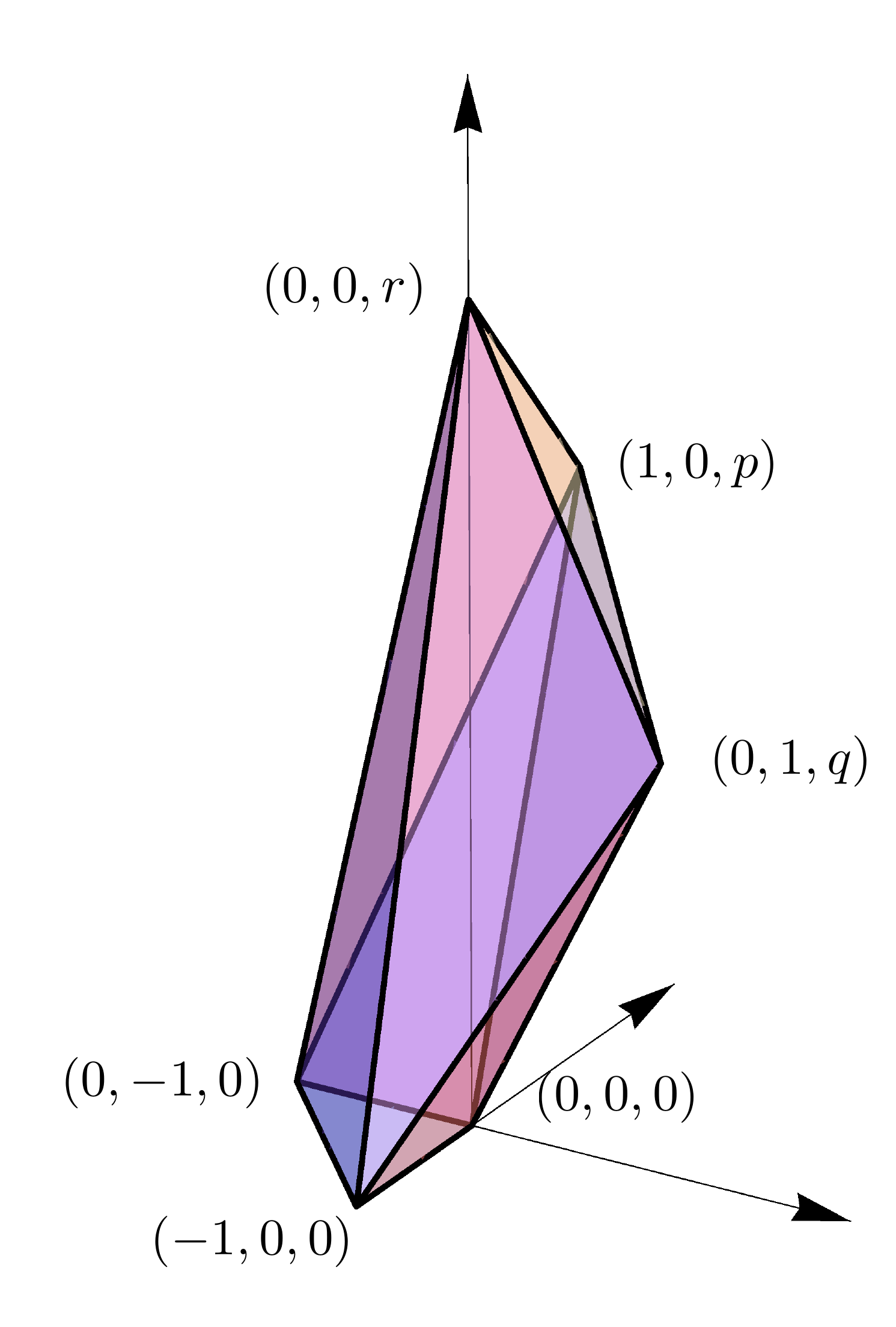}
    \caption{The toric diagram for $A^{p,q,r}$. For $p=q$ we obtain the manifolds $Y^{r,p}\left ( \cc\mathbb{P}^1\times \cc\mathbb{P}^1\right )$.}
    \label{fig:td}
\end{figure}
\begin{equation}\hspace{-.5cm}
	\begin{array}{cccc}
		[p,0,0,-1]\ ,\quad & 
		[p,q,0,-1]\ ,\quad & 
		[0,0,0,-1]\ ,\quad &
		[0,q,0,-1] \\
		{[-p+r,-r,-r,1]}\ ,\quad & 
		[-p+r,-q+r,-r,1]\ ,\quad &
		[-r,-r,-r,1]\ ,\quad & 
		[-r,-q+r,-r,1]\ .
	\end{array}
\end{equation}
The dual cone, the fan of the toric manifold, is specified by the following vectors, orthogonal to the six facets of the polyhedron and pointing outwards: 
\begin{equation}
	\begin{array}{ccc}
	  [0,0,1,0]\ , \qquad & [0,0,1,r]\ , \qquad &  [-1,0,1,0] \\
	  {[1,0,1,p]}\ ,  \qquad & [0,-1,1,0]\ , \qquad & [0,1,1,q]\ .
	\end{array}
\end{equation}
All these vectors lie on a hyperplane, as required by the Calabi--Yau condition. The projection of the fan on the common hyperplane is the toric diagram,  pictured in figure \ref{fig:td}. For $p=q$ we recover the known toric diagram of $Y^{r,p}\left (\cc\mathbb{P}^2\times \cc\mathbb{P}^1\right )$ \cite{martelli-sparks-sasaki7}. 

One can check from the form of the toric diagram that the $A^{p,q,r}$ are a subset of a more general family of Sasaki--Einstein manifolds with cohomogeneity two and ${\rm U}(1)^4$ symmetry discussed in \cite{lu-pope-vazquezporitz2} in different coordinates; the family depends on six integers and has a toric diagram with six external points. It should be possible, although not obvious, to find a suitable change of coordinates for (a subset of) the metrics in \cite{lu-pope-vazquezporitz2} that makes the ${\rm SU}(2)\times {\rm SU}(2)$ invariance manifest.




\section{Massive solutions} 
\label{sec:massive}

We will now examine the solutions of the system (\ref{eq:sys}) with $F_0\neq 0$. 

We have not been able to find analytical solutions; we have analyzed the system numerically. Fortunately, the regularity of the solutions was analyzed in section \ref{sub:global} irrespectively of the details of the solution. These numerical results should be robust: some of the solutions presented here are a $\zz_2$ quotient of the ones examined in \cite{ajtz}, which matched predictions from AdS/CFT with great accuracy. 

We used as initial conditions the local solution (\ref{eq:equator}) around the equator $t=\pi/4$. Namely, we took an initial $t_0= \pi/2 + \epsilon$, and we chose the initial values for $\psi$ and $w_i$ according to (\ref{eq:equator}) (neglecting the $O((t-\pi/2)^3)$ terms because of the smallness of $\epsilon$). We did this for both positive and negative $\epsilon$, obtaining the solution in the North and South hemisphere ($t<\pi/4$ and $t>\pi/4$). 

\begin{figure}[h]
    \centering
        \includegraphics[width=29em]{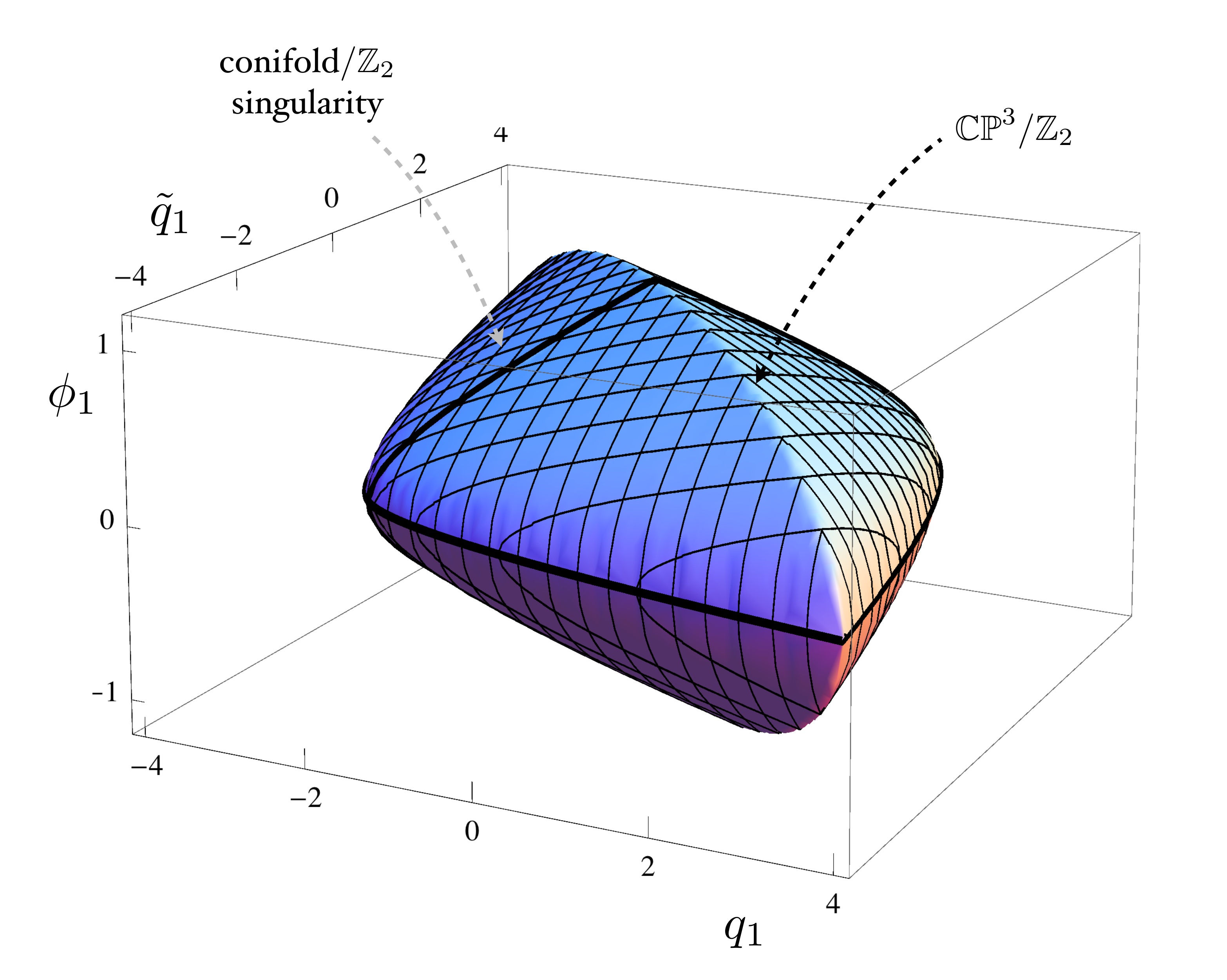}
    \caption{The allowed values for the parameters $(q_1,\tilde q_1, \phi_1)$ are the ones in the interior of this diagram. The intersection of this picture with the $\phi_1$ plane is the diagram in figure \ref{fig:ab}. A generic point on the boundary represents a manifold with a local $\zz_2$ singularity. Points on the thick black line represent manifolds with a conifold$/\zz_2$ singularity. The ``ridge'' at $q_1=\tilde q_1$ represents spaces with the topology of $\cc\pp^3/\zz_2$. The point at the very top, which lies at the intersection of the ridge and the black line, represents a space that has two conifold$/\zz_2$ singularities.}
    \label{fig:qqp}
\end{figure}

This procedure can be repeated for all $(q_1,\tilde q_1, \phi_1)$. We show in figure \ref{fig:qqp} the values of the parameters for which the solution exists for all $t\in [0, \pi/2]$.

This parameter space is symmetric under three reflections: 
\begin{equation}
	(q_1,\tilde q_1, \phi_1) \to (\tilde q_1,q_1, \phi_1) \ ,\qquad
	(q_1,\tilde q_1, \phi_1) \to (-\tilde q_1,-q_1, \phi_1) \ ,\qquad
	(q_1,\tilde q_1, \phi_1) \to (q_1, \tilde q_1, -\phi_1) \ .
\end{equation}
All three of them are explained by a symmetry of the system: the first by (\ref{eq:symw}), the second by (\ref{eq:symw}) combined with (\ref{eq:symt}), the third by (\ref{eq:sympsi}).

The intersection of this diagram with the plane $\phi_1=0$ is nothing but the diagram in figure \ref{fig:ab}. For points in the interior of the three--dimensional diagram in figure \ref{fig:qqp}, $w_i \neq 0$ everywhere. In this case, we can apply the local analysis in (\ref{eq:NPgen}) to conclude that the solution is non--singular.

For generic points on the boundary of the diagram, one of the two $w_i$ vanishes (i.e.~one of the two $S^2$ shrinks) at either $t=0$ or $t=\pi/2$. For these points, we can apply the local analysis in (\ref{eq:NPsp}) to see that the corresponding manifolds have a $\zz_2$ orbifold singularity.

Let us divide the boundary of the diagram in figure \ref{fig:qqp} in four vertical meridians: $-q_1<\tilde q_1<q_1$, $-\tilde q_1 < q_1 < \tilde q_1 $, $q_1 < \tilde q_1 < -q_1$ and $\tilde q_1 < q_1 < -\tilde q_1$. These are bounded by the vertical ``ridge'' and by the vertical black line. The intersection of each of these four meridians with the $\phi_1=0$ plane is then one of the four components of the boundary of the diagram in figure \ref{fig:ab}. This dictates which sphere shrinks on which meridian. For example, points in the meridian $-q_1 < \tilde q_1 < q_1$ correspond to solutions for which the sphere $S^2_1$ shrinks at $t=0$, just as in figure \ref{fig:ab}. 
  
The points on the vertical ``ridge'' correspond to manifolds whose topology is $\cc\pp^3/\zz_2$, just like the points $(\pm 4, \mp 4)$ in figure \ref{fig:ab} (which are indeed the intersections of the ridge with the plane $\phi_1=0$). Points on the black vertical line on the boundary correspond to points where two $S^2$ shrink in the same point, just like the points $(\pm \frac 4{\sqrt{3}}, \pm \frac 4 {\sqrt{3}})$ (which are indeed the intersections of the black line with the plane $\phi_1=0$). In this case, we can apply the local analysis in (\ref{eq:NPcon}) to learn that the space has a local conifold$/\zz_2$ singularity. Finally, for the point at the very top of the diagram in figure \ref{fig:qqp}, both $S^2$ shrink on both sides. This space has two conifold$/\zz_2$ singularities.

The parameter space we have obtained is expected to be discretized by the flux quantization conditions, but densely, just like for the parameter space of massless solutions in figure \ref{fig:ab} (see the comment after (\ref{eq:pi3/pii})).\footnote{There do exist, however, moduli that are not visible in our Ansatz, as we will see in section \ref{sub:moduli}.} This is similar to what was found for ${\cal N}=1$ solutions in \cite{t-cp3,koerber-lust-tsimpis}. An important difference is that the boundary of the parameter space of solutions in this paper (and the ones in \cite{ajtz}) represents singular solutions, whereas the boundary of the parameter space of  ${\cal N}=1$ solutions in \cite{t-cp3,koerber-lust-tsimpis} represents massless solutions.

The parameter space in figure \ref{fig:qqp} also has a vague similarity with the chambers which tessellate the K\"ahler moduli space of Calabi--Yau manifolds. Recall that each of these chambers is bounded by ``walls'' where some two--cycle shrinks; one can go beyond these walls, however, and end up in another chamber, by performing a flop. There are also ``extremal transitions'' where one or more shrunk three--cycles are replaced by two--cycles. (For a review of these phenomena, see for example \cite{greene-review}). 

It would be interesting to know whether there is anything similar in our case. Unfortunately, most of the mathematical techniques that helped in the Calabi--Yau case cannot be adapted easily. Since our spaces are not K\"ahler (and not even complex), for example, algebraic geometry cannot be applied. 

One point of contact with the Calabi--Yau case might be the the emergence of light branes, which is the physics way of ``predicting'' extremal transitions \cite{strominger-conifold}. For the solutions analyzed in \cite{ajtz}, light branes do indeed appear. The parameter space in that case is a segment, describing the allowed range for a parameter $\psi_1$ similar to $\phi_1$ in this paper. A $\zz_2$ symmetry (essentially the same as in (\ref{eq:sympsi})) maps $\psi_1 \to - \psi_1$; the case $\psi_1=0$ corresponds to the massless ${\cal N}=6$ solution on AdS$_4\times \cc\pp^3$, whereas each of the endpoints of the segment corresponds to a space with a conifold singularity.   In terms of the dual gauge theory parameters, the singular locus corresponds to the large $N$ limit at fixed Chern--Simons levels. There are light states, coming from supersymmetric D2/D0 branes wrapped on a collapsing cycle, which are dual to monopoles operators. Their existence and dimension can be predicted by purely field--theoretic arguments \cite{ajtz}.

Since the solutions in \cite{ajtz} are essentially a $\zz_2$ cover of the solutions on the ``ridge'' in figure \ref{fig:qqp}, one expects some similar phenomena for the solutions discussed in this paper; it would be interesting to perform a detailed analysis of this. A dual Chern--Simons gauge theory covering at least part of the parameter space will be proposed in section \ref{sec:quiver}. We leave for future work the identification of the singular loci in the parameter space in terms of gauge theory parameters and the study of the spectrum of light BPS operators. 


\section{A dual Chern-Simon quiver} 
\label{sec:quiver}

It is interesting to identify the CFT$_3$ dual to the M--theory and type IIA backgrounds that we have discussed in this paper. We will first identify a possible dual for the M--theory solution AdS$_4\times A^{p,q,r}$, at least for some values of the integers $p,q,r$, and then we move on to their massive deformations.  In subsection \ref{sub:moduli} we use this quiver to count the number of moduli of our solutions.

\subsection{The quiver} 
\label{sub:quiver}

To propose a theory dual to the massless solutions, we will look for an ${\cal N}=2$  quiver with the expected global symmetries, and whose moduli space of vacua reproduces the spaces $A^{p,q,r}$. This strategy has given satisfactory results in the case of D3 branes at toric singularities and has been recently applied to membranes theories \cite{martelli-sparks-3dquivers,hanany-zaffaroni-cs, hanany-vegh-zaffaroni}. We should note that
the evidence here is not as strong as in  the D3 brane case, where many comparisons can be done with the quantum field theory predictions. The indirect evidence of this kind of general strategy  is also weaker than the chain of string dualities that establish the holographic duals of the ${\cal N}=6$ \cite{abjm} and ${\cal N}=3$ \cite{jafferis-t} theories
\footnote{Attempts to reconstruct ${\cal N}=2$  quivers  from a string construction have been made in \cite{aganagic}, and in \cite{Benini:2009qs,Jafferis:2009th} for theories with fundamental matter.}. However, the  quiver we propose is a natural generalization and a $\mathbb{Z}_2$ quotient of the one used in \cite{ajtz}, which passed quite non-trivial checks.

\begin{figure}[h]
    \centering
        \includegraphics[width=29em]{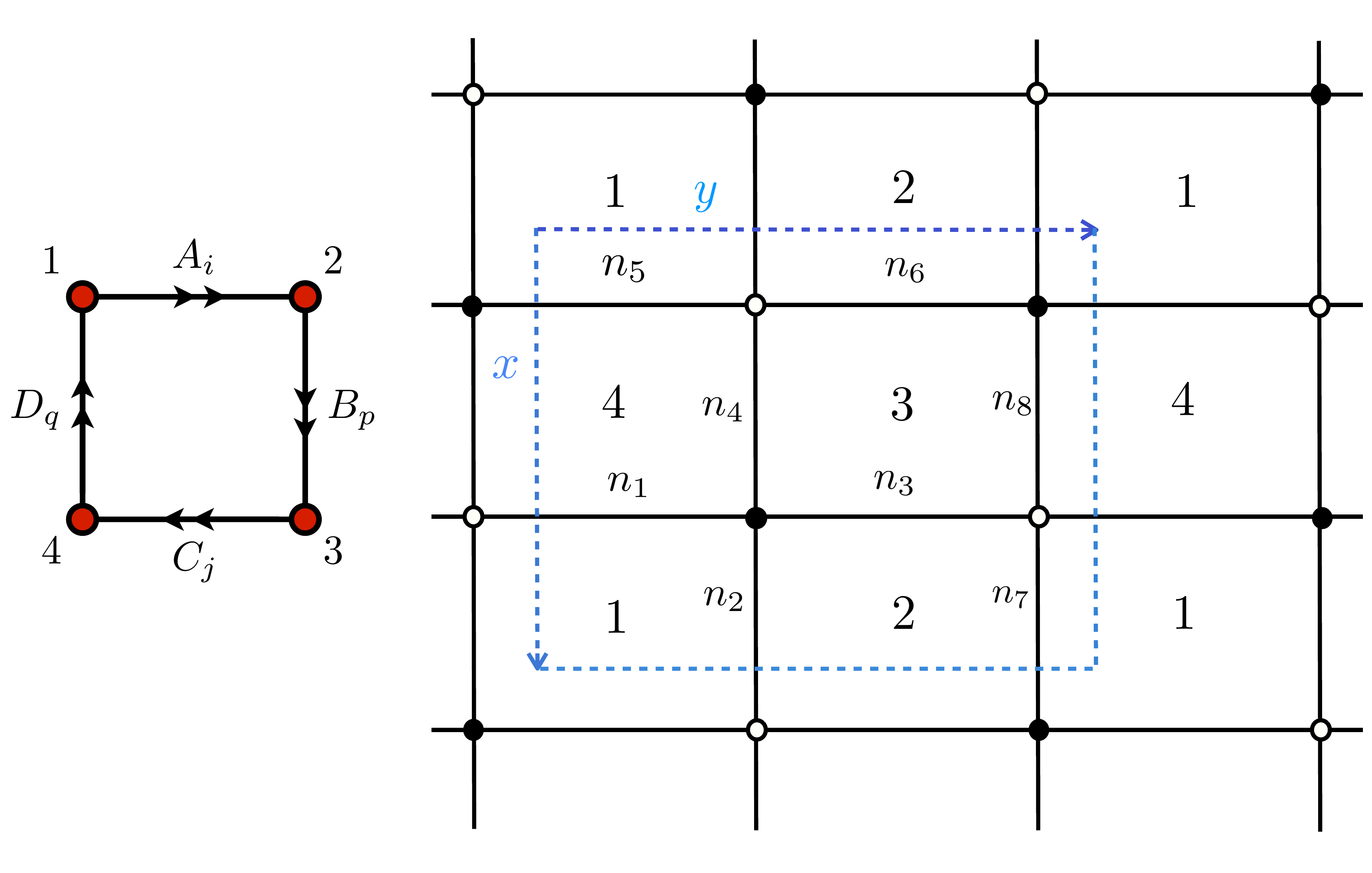}
    \caption{The quiver and Tiling for $A^{p,q,r}$. }
    \label{fig:tal}
\end{figure}

With this caveat in mind, consider the quiver pictured in figure \ref{fig:tal}. It has gauge group ${\rm U}(N)^4$, and eight chiral bi--fundamental
fields $A_i,\, B_i,\, C_i,\, D_i$, $i=1,2$ transforming in the representation $(N,\,  \bar{N},\, 0,\, 0), \,$
$(0, N,\,  \bar{N},\, 0),$ $(0,\, 0,\, N,\,  \bar{N}), $ $(\bar{N},\, 0,\, 0,\, N)$ of the gauge group, respectively, and interacting with the superpotential
\begin{equation}\label{eq:W}
W=\epsilon_{ij}\epsilon_{pq} A_i\, B_p\, C_j\, D_q \ .
\end{equation}
The theory has a global ${\rm SU}(2)\times {\rm SU}(2) \times {\rm U}(1)_{\rm R}$ symmetry that reflects the non-abelian isometry of $A^{p,q,r}$.
In the context of four--dimensional CFTs, the quiver in figure \ref{fig:tal} is used to describe D3 branes at the $\mathbb{F}_0$ singularity.
As usual in applications to three--dimensional CFTs, we introduce no kinetic term for the gauge fields but a Chern--Simons interaction with coefficient $k_i$ with $\sum_{i=1}^4 k_i=0$. The resulting theory will describe membranes probing Calabi--Yau four--folds  \cite{jafferis-t,martelli-sparks-3dquivers,hanany-zaffaroni-cs}. In order to find the right dual, we need to find an  ${\cal N}=2$ Chern--Simons theory whose moduli space is the Calabi--Yau cone over $A^{p,q,r}$. For specific choices of the Chern-Simons parameters, this quiver has been shown to  describe $Q^{1,1,1}$ and its quotients \cite{davey-hanany-mekareeya-torri,amariti-forcella-girardello-mariotti,aganagic}. It was suggested in \cite{martelli-sparks-3dquivers} that it might describe in general the $Y^{p,q}\left(\mathbb{CP}^1\times \mathbb{CP}^1\right )$ manifolds. We will see now that it actually describes part of the family $A^{p,q,r}$.
    
\begin{figure}[h]
    \centering
        \includegraphics[width=30em]{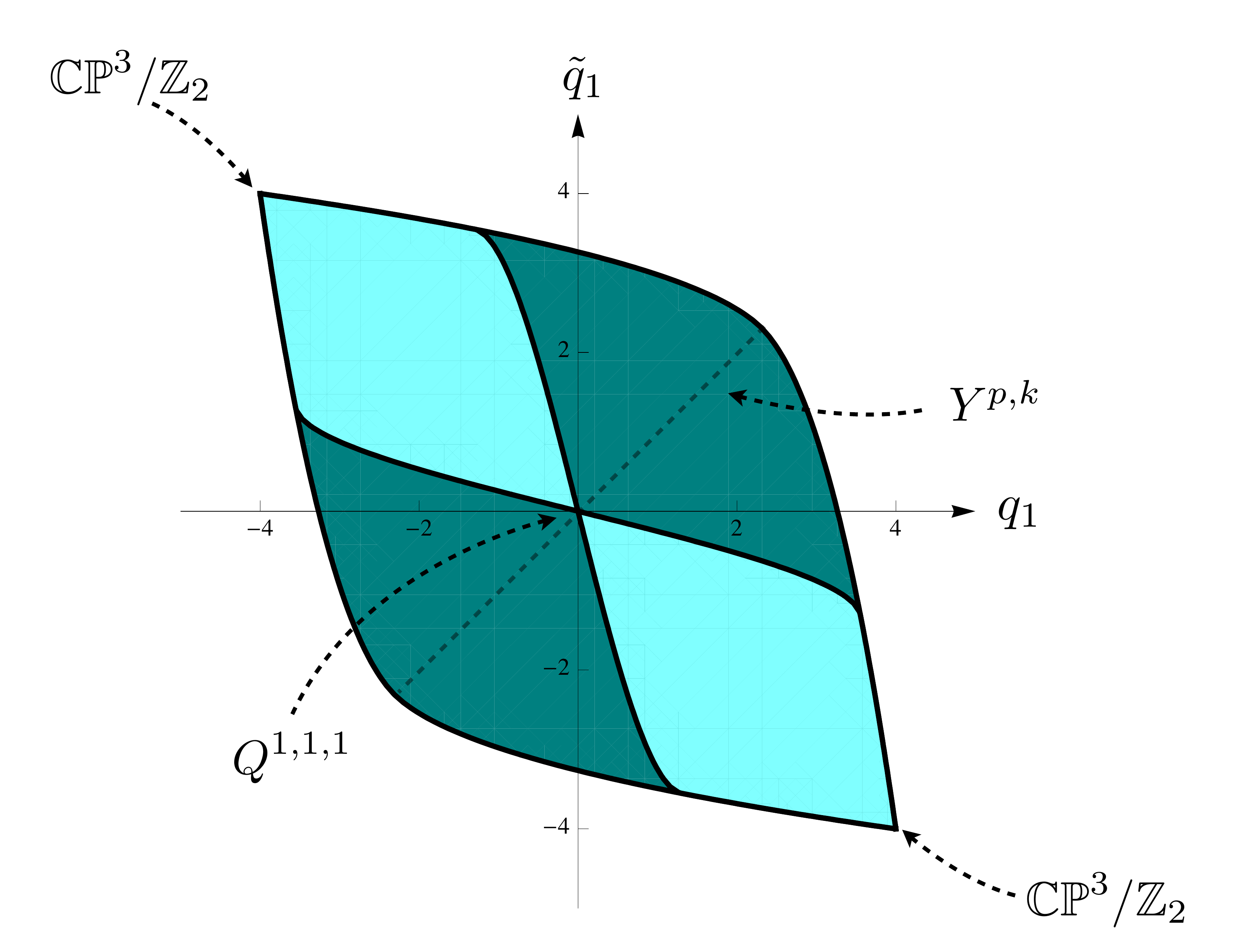}
    \caption{The light region corresponds to the Sasaki--Einstein manifolds whose dual quiver is the one in figure \ref{fig:tal}.}
    \label{fig:ab2}
\end{figure}

We can compute the moduli space of the quiver using its Tiling realization and the associated Kasteleyn matrix\cite{hanany-zaffaroni-cs, hanany-vegh-zaffaroni}. In this language, every face in the Tiling is a gauge group and every link is a chiral bi--fundamental with orientation determined by the position of the white and black nodes. We compute the moduli space with the algorithm discussed in section (2.1.1) of \cite{davey-hanany-mekareeya-torri}, to which we refer for details.  We introduce an  integer number $n_a$ for each  link in the Tiling in such a way that 
\begin{equation}
\sum_{a\, \in \, i{\rm -th\,  face} } d^{a}_{i} n_a =k_i\ ,
\end{equation}
where $d^{a}_{i}=\pm 1$ is the ${\rm U}(1)$ charge of the field corresponding to the link \cite{hanany-zaffaroni-cs,imamura-kimura-crystals,davey-hanany-mekareeya-torri}. Different choices of the $n_a$ give equivalent results; we can set $n_1=n_3=n_7=k_4$, $n_2=k_1$,  $n_6=k_1+k_2$, $n_4=n_5=n_8=0$. Recall that $\sum_{i=1}^4 k_i=0$ and therefore $k_3=-k_1-k_2-k_4$. The modified Kasteleyn matrix is easily computed to be:
\begin{equation}
\left( 
\begin{array}{cc}
B_2 z^{n_6} + D_1  \frac{z^{n_5}}{x}\, ,  & C_1 z^{n_8} + A_2 \frac{z^{n_7}}{y} \\
C_2 z^{n_4} + A_1 z^{n_2} y\, , & B_1 z^{n_3} + D_2 z^{n_1} x
\end{array} \right ) \ ,
\end{equation}
and its determinant (with all the fields set to $1$) reads
\begin{equation} 
	1 + \frac{1}{x}-\frac{1}{y}- z^{k_1} +z^{k_1+k_2}+ x z^{k_1+k_2}- y z^{k_1-k_4}- z^{-k_4}\ .
\end{equation}
From this Laurent polynomial we read the points of the toric diagram:
\begin{equation}
	\begin{array}{cccc}
		[0,0,0]\ ,\qquad &[-1,0,0]\ ,\qquad & [0,-1,0]\ ,\qquad &
		[0,0,k_1]\ ,\\
		{[0,0,k_1+k_2]}\ ,\qquad &
		[1,0,k_1+k_2]\ ,\qquad &
		[0,1,k_1-k_4]\ ,\qquad &
		[0,0,-k_4]\ .
	\end{array}
\end{equation}
Two of the points in this diagram are necessarily internal. 
There are various ways  to reproduce the diagram  in figure \ref{fig:td}.   For example,  we can require that  the two  points $ [0,0,k_1+k_2]\, , [0,0,-k_4]$ be internal and  choose $k_1=r$, $k_2=p-r$, $-k_4=q-r$. 
The corresponding quiver has Chern-Simons parameters $(r,p-r,r-q,q-p-r)$.
Consistency  requires  $r\le q\le 2 r$ and $0\le p\le r$.  Other  choices of internal points  give equivalent results\footnote{Notice that there are other choices with points with negative third coordinate. One of these models correspond for example again to $Q^{1,1,1}/\mathbb{Z}_2$ \cite{davey-hanany-mekareeya-torri,amariti-forcella-girardello-mariotti,klebanov}.}.
Obviously we could also obtain the symmetric case with the role of $p$ and $q$ interchanged.
The region in parameter space covered by the quiver is the light one in figure \ref{fig:ab2}. 
We see that we can recover all $A^{p,q,r}$ manifolds with $r\ge p$ and $r\le q$. In particular we see that the manifolds $Y^{r,p}= A^{p,p,r}$ are represented only in the case $r=p$, which corresponds to a $\zz_p$ quotient of $Q^{1,1,1}$; the corresponding quiver, with Chern--Simons parameters $(p,0,-p,0)$, was discussed in \cite{aganagic}. Another limiting case is $p=0$ and $q=2 r$, which correspond to the corner $\mathbb{CP}^3/\mathbb{Z}_2$, whose lift to M theory is 
a quotient of $S^7$; the corresponding quiver, with Chern--Simons parameters $(r,-r,r,-r)$, was discussed in \cite{hanany-zaffaroni-cs} and shown indeed to correspond to quotients of the Calabi--Yau cone $\mathbb{C}^2/\mathbb{Z}_2\times \mathbb{C}^2/\mathbb{Z}_2$.

Up to now, we considered a quiver with equal gauge groups and Chern--Simons parameters adding to zero. There are four integer parameters, namely $N$ and $k_1$, $k_2$, $k_4$, which correspond to the number of membranes and the three integers $p,q,r$ defining the Sasaki--Einstein manifold. In the type IIA picture, these integers correspond to the possible values of the $F_2$ and $F_6$ fluxes.  The massive type IIA solutions discussed in this paper have a total of eight integer parameters, the RR fluxes on zero--, two--, four-- and six--cycles. The Romans mass $F_0$ is dual to the sum of the Chern--Simons parameters $\sum_{i=1}^4 k_i$   \cite{gaiotto-t,fujita-li-ryu-takayanagi}, while the differences in ranks are dual to the four--form flux \cite{aharony-bergman-jafferis}.  It is then natural to conjecture that the solutions describe the quiver Chern--Simons theory for generic values of the eight parameters $N_i$ and $k_i$, $i=1,\ldots,4$ as in \cite{gaiotto-t2,petrini-zaffaroni,lust-tsimpis-singlet-2,ajtz}.  We note that the ``ridge'' in figure \ref{fig:qqp} is simply a $\mathbb{Z}_2$ quotient of the solution in \cite{ajtz} describing the ${\rm SU}(2)\times {\rm SU}(2)$ invariant ${\cal N}=2 $ deformation of the ABJM theory.
 
 It would be interesting to consider also the second toric phase of the $\mathbb{F}_0$ quiver as in \cite{hanany-vegh-zaffaroni} and other examples of quivers for $Q^{1,1,1}$ \cite{franco-hanany-park-rodriguezgomez} with arbitrary Chern--Simons couplings, to see if we can obtain a more complete or complementary description of $A^{p,q,r}$ and of the massive solution. We leave this for future work.


\subsection{An application: moduli} 
\label{sub:moduli}

We will now count the number of moduli of our solutions, by computing the dimension of the conformal manifold of the theory. We will perform that computation as in \cite{green-komargodski-seiberg-tachikawa-wecht}. Namely, we first count the number of marginal operators, and we then quotient it by the complexification of the group of global symmetries. 

As a warm--up, let us consider the massive solutions in \cite{ajtz}. The quiver proposed there is the same as the ABJM quiver \cite{abjm}, although, as we recalled above, the presence of $F_0$ requires the Chern--Simons levels not to sum up to zero \cite{gaiotto-t}. The proposed quiver matches the prediction from the AdS/CFT correspondence with great accuracy  \cite{ajtz}. The fields $A_i$, $B_i$ have dimension $1/2$ as in \cite{abjm}. The most general operator with the right dimension is then of the form 
\begin{equation}\label{eq:marginal}
	O^{ijkl}{\rm Tr}(A_i B_k A_j B_l)\ ;
\end{equation}
the F--term equations, however, demand that $B_{[k} A_j B_{l]}=0$, $A_{[i} B_k A_{j]}=0$, so that the tensor $O^{ij,kl}$ can actually be taken to be symmetric: 
\begin{equation}\label{eq:Osym}
	O^{ijkl}= O^{jikl}=O^{ijlk}\ .
\end{equation}
If we think of the first and second pair of indices as a single index each, $O$ represents then a $3\times 3$ matrix: it has 9 complex entries. The flavor group is ${\rm SU}(2)\times {\rm SU}(2)$, whose complexification has complex dimension 6. This leaves us with a conformal manifold of dimension 3. 

This tells us that the solutions in \cite{ajtz} have 3 complex moduli. In that paper, we had as many parameters as we had flux quantization conditions, leaving us with no continuous parameters. However, the Ansatz used there was one compatible with ${\rm SU}(2)\times {\rm SU}(2)$ symmetry, whereas the 3 moduli we just found break this symmetry by construction. So there is no contradiction. It would be interesting to find the three parameter family of
massless and massive type IIA  solutions with $AdS_4$ vacuum corresponding to this moduli space.

In a sense, the 3 parameters should correspond to an $O$ which is antisymmetric under exchange of the first and second pair of indices ($O^{ijkl}=-O^{klij}$). This operation corresponds roughly to an exchange of the two $S^2$s in the metric; one could imagine getting rid of these moduli by some appropriate orbifold. Even if this worked, however, this would eliminate flat directions; in principle one would still be left with the six marginal but not exactly marginal operators, which should correspond holographically to scalars which are massless but whose potential is not flat. Such scalars are as unpleasant as proper moduli, from the point of view of any application to four--dimensional compactifications. 

We now leave the solutions in \cite{ajtz} and turn to the ones discussed in this paper. Actually, the counting proceeds in a very similar fashion. We do not know the dimensions of the fields, but from the superpotential (\ref{eq:W}) we know that the sum of the dimensions of $A$, $B$, $C$, $D$ has to be 2. Hence,
marginal operators have the form 
\begin{equation}
	O^{ijkl}{\rm Tr}(A_i B_k C_j D_l)\ ,
\end{equation}
similar to (\ref{eq:marginal}). The F--term equations enforce again (\ref{eq:Osym}), which means that we have again 9 marginal operators. Since the flavor symmetry is again ${\rm SU}(2)\times {\rm SU}(2)$, we again end up with a conformal manifold of dimension 3. We stress again that this prediction is based on the proposed duality
between the quiver and the supergravity solution.

It would be interesting to find quivers which have no exactly marginal operators (or better still, no marginal operators to begin with). If such quivers have gravity duals, these would be supergravity vacua with no moduli (albeit with negative cosmological constant). This situation would be comparable to the vacua in \cite{kklt} before the last ``uplifting'' stage, or to the solutions in \cite{dewolfe-giryavets-kachru-taylor}.



\section*{Acknowledgments}
We wish to thank S.~Cremonesi and D.~Martelli for interesting discussions.
 We are supported in part by INFN and MIUR under contract 2007--5ATT78--002.


\providecommand{\href}[2]{#2}

\end{document}